\documentclass[prc,preprint,
  showpacs,showkeys,lengthcheck,
  nofootinbib,tightenlines,onecolumn,notitlepage,
  preprintnumbers,superscriptaddress]{revtex4-1}

\usepackage{graphicx}
\usepackage[usenames,dvipsnames]{xcolor}
\graphicspath{{figures/}}

\usepackage{xr-hyper}

\usepackage{hyperref}
\hypersetup{colorlinks=true,citecolor=blue,linkcolor=blue,urlcolor=blue}

\usepackage{diagbox}
\usepackage{booktabs}

\usepackage{amsmath,amssymb,amsfonts}
\usepackage{mathrsfs}
\usepackage{slashed}
\usepackage{bm}

\begin{document}

\title{Gravitational form factors of light mesons}

\author{Adam Freese}
\email{afreese@anl.gov}
\affiliation{Argonne National Laboratory, Lemont, Illinois 60439, USA}

\author{Ian C. Clo\"{e}t}
\email{icloet@anl.gov}
\affiliation{Argonne National Laboratory, Lemont, Illinois 60439, USA}

\begin{abstract}
  We calculate the gravitational form factors of
  the pion, sigma meson, and rho meson
  in the Nambu--Jona-Lasinio (NJL) model of quantum chromodynamics.
  The canonical energy-momentum tensor (EMT) is used in their derivation,
  allowing the possibility of an antisymmetric
  contribution when the hadron has intrinsic spin.
  We show that the asymmetric graviton vertex arising from
  the canonical EMT satisfies a simpler Ward-Takahashi identity than
  the symmetric graviton vertex of the Belinfante EMT.
  The necessity of fully dressing the graviton vertex through
  the relevant Bethe-Salpeter equation is demonstrated
  for observing both the WTI and a low-energy pion theorem.
  Lastly, we calculate static moments of the meson EMT decompositions,
  obtaining predictions for the meson mass radii.
  We find light cone mass radii of 0.27~fm for the pion,
  0.32~fm for the sigma, and 0.25~fm for the rho.
  For the pion and rho, these are smaller than the light cone charge radii,
  respectively 0.51~fm and 0.45~fm,
  while we have a sigma charge radius of zero.
  Our light cone pion mass radius agrees with
  a phenomenological extraction from KEKB data.
\end{abstract}

\maketitle

\onecolumngrid


\section{Introduction}
\label{sec:intro}

In recent years, the energy-momentum tensor (EMT) of hadrons
has become an increasingly popular object of study
in our quest to better understand hadronic structure---and,
in turn, quantum chromodynamics (QCD) itself.
(See Refs. \cite{Leader:2013jra,Polyakov:2018zvc} and references therein.)
Understanding the EMT can help address such fundamental questions as
where does the mass of the proton come from, and where is all of its spin?
It also opens new avenues for exploration,
including not just spatial distributions of energy, momentum, and angular momentum
\cite{Polyakov:2002yz,Leader:2013jra,Lorce:2017xzd,Lorce:2017wkb},
but also the distribution of pressure and shear forces
\cite{Polyakov:2002wz,Polyakov:2002yz,Polyakov:2018zvc,Lorce:2018egm}.

It has been remarked that the EMT introduces a new intrinsic global quantity
in addition to mass and spin,
called the ``D-term'' \cite{Polyakov:2002wz,Polyakov:2018zvc}.
The D-term, which quantifies the strength of the forces binding the
hadron together, is not constrained by conservation laws nor
by representation theory of the Lorentz group.
However, in certain cases it may be constrained by other considerations---as
it is with Nambu-Goldstone bosons \cite{Novikov:1980fa,Voloshin:1980zf,Polyakov:1999gs}.
Since the pion plays a vital role in QCD as the Nambu-Goldstone boson
of dynamical chiral symmetry breaking (DCSB) \cite{Nambu:1961tp,Nambu:1961fr},
understanding how its D-term comes to be constrained may play as important a
role in fully grasping QCD as understanding the origin of proton mass and spin.

DCSB has come to be understood as one of the central features of QCD,
and is intimately involved in the observed mass of hadrons
\cite{Bashir:2012fs}.
Accordingly, understanding the pion is vital to understanding QCD,
and pion structure has long been a major topic of study in hadron physics.
Its EMT in particular has recently been computed on the lattice
\cite{Brommel:2007zz,Shanahan:2018pib}.
Additionally, it has become possible through dispersive analysis of KEKB data
for the reaction $\gamma^*\gamma\rightarrow\pi^0\pi^0$
to extract an empirical parametrization of the pion EMT
in the timelike region \cite{Kumano:2017lhr}.
The prospect of comparison to empirical values makes model calculations
of the pion EMT especially important now.

In this work, we investigate the EMT of the pion and of other mesons
in the Nambu--Jona-Lasinio (NJL) model
\cite{Vogl:1991qt,Klevansky:1992qe,Hatsuda:1994pi} of QCD.
The NJL model is a Poincar\'{e} covariant quantum field theory
that successfully reproduces low-energy properties of QCD such as DCSB.
Additionally, since the model does not contain gluons,
we can defer issues regarding gauge invariance of the EMT.
These properties make the NJL model an ideal framework in which to investigate
the EMT of the pion, especially aspects of the EMT that arise from DCSB.
The sigma and rho mesons are also investigated as a point of contrast,
with the latter serving to illustrate the ways in which spin manifests
in the EMT.

This work is organized into the following sections.
We first give an overview of the NJL model in Sec.~\ref{sec:NJL}.
We then develop the formalism needed to calculate the meson EMTs
in Sec.~\ref{sec:graviton},
where the formalism is cast as a study of quark-graviton
and meson-graviton interactions.
In Sec.~\ref{sec:axial},
we develop the formalism needed to calculate the axial form factors
of the rho meson, which are needed to numerically demonstrate
a correspondence between certain gravitational and axial form factors.
Results for the meson EMTs are given in Sec.~\ref{sec:results},
where we also explore implications of our results.
Finally, we conclude in Sec.~\ref{sec:concl} and give a brief outlook,
including the prospects of performing similar calculations in QCD.


\section{The NJL model}
\label{sec:NJL}

The Nambu--Jona-Lasinio (NJL) model, originally proposed as a theory
of elementary nucleons \cite{Nambu:1961tp,Nambu:1961fr},
is used as a low-energy effective field theory
that models the dynamical chiral symmetry breaking of quantum chromodynamics
through a four-fermi contact interaction
\cite{Vogl:1991qt,Klevansky:1992qe,Hatsuda:1994pi}.
It has successfully been applied to modeling the physical properties
of both mesons \cite{Vogl:1991qt,Klevansky:1992qe,Cloet:2014rja,Ninomiya:2017ggn}
and baryons \cite{Ishii:1993np,Ishii:1993rt,Ishii:1995bu,Cloet:2014rja}.
Accordingly, we employ the NJL model in this work to calculate
the energy-momentum tensor of light mesons.

The two-flavor NJL model Lagrangian we use is \cite{Cloet:2014rja}:
\begin{align}
  \mathcal{L}
  &=
  \overline{\psi}( i \overleftrightarrow{\slashed{\partial}}-\hat{m}) \psi
  + \frac{1}{2} G_\pi [
    (\overline\psi\psi)^2 - (\overline{\psi}\gamma_5\boldsymbol{\tau}\psi)^2
    ]
  - \frac{1}{2} G_\omega (\overline\psi\gamma_\mu\psi)^2
  - \frac{1}{2} G_\rho [
    (\overline\psi\gamma_\mu\boldsymbol{\tau}\psi)^2
    + (\overline{\psi}\gamma_\mu\gamma_5\boldsymbol{\tau}\psi)^2
    ]
  \notag \\
  &+ \frac{1}{2} G_\eta [
    (\overline\psi\boldsymbol{\tau}\psi)^2 - (\overline{\psi}\gamma_5\psi)^2
    ]
  - \frac{1}{2} G_f (\overline{\psi}\gamma_\mu\gamma_5\psi)^2
  - \frac{1}{2} G_T [
    (\overline\psi i\sigma^{\mu\nu}\psi)^2
    - (\overline{\psi}i\sigma^{\mu\nu}\boldsymbol{\tau}\psi)^2
    ]
  \label{eqn:lagrangian}
  \,,
\end{align}
where $\hat{m} = \mathrm{diag}[m_u,m_d]$ is the current quark mass
matrix (where we take $m_u=m_d\equiv m$ in this work),
$\tau_i$ are the isospin matrices, and
the $G_i$ are four-fermi coupling constants.
The expression in Eq.~(\ref{eqn:lagrangian}) is invariant under
$\mathrm{SU}(2)_V$ and $\mathrm{SU}(2)_A$ transformations,
but $\mathrm{U}(1)_A$ symmetry requires the additional constraints
$G_\eta = G_\pi$ and $G_T = 0$, which we will assume in this work.
We take the Lagrangian in Eq.~(\ref{eqn:lagrangian}) to symbolically
represent a Lagrangian that has already been Fierz symmetrized,
so that only direct terms need to be calculated in the interaction kernel
(See Ref. \cite{Ishii:1995bu} for a detailed description of the Fierz
symmetrization procedure.)

The NJL model dynamically generates a dressed quark mass $M$
when $G_\pi > G_{\mathrm{critical}}$.
This dynamical mass generation is described by the gap equation:
\begin{align}
  M = m + 2iG_\pi (2N_c)
  \int\frac{\mathrm{d}^4k}{(2\pi)^4}
  \mathrm{Tr}_D\left[ S(k) \right]
  =
  m
  + 8i G_\pi (2N_c)
  \int\frac{\mathrm{d}^4k}{(2\pi)^4}
  \frac{M}{k^2-M^2+i0}
  \label{eqn:gap}
  \,,
\end{align}
where the trace is over the Clifford matrix structure
(with $2N_c$ having come from the color and isospin traces already).

\begin{figure}
  \centering
  \includegraphics[width=0.6\textwidth]{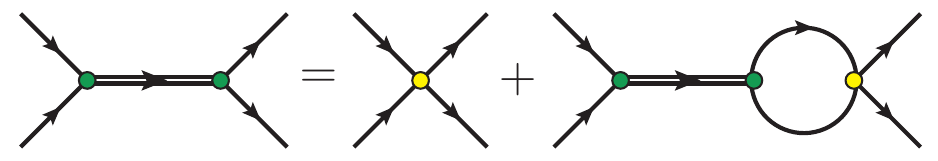}
  \caption{
    Diagrammatic depiction of the inhomogeneous Bethe-Salpeter
    equation for the two-body T-matrix in the NJL model.
  }
  \label{fig:tmatrix:bse}
\end{figure}

Mesons appear in the NJL model as bound state poles in the
quark-antiquark T-matrix,
which itself can be found from solving an inhomogeneous
Bethe-Salpeter equation,
depicted diagrammatically in Fig.~\ref{fig:tmatrix:bse}.
Poles corresponding to mesons with the quantum numbers of various mesons
are present in the NJL T-matrix.
Our interest in this work lies primarily with the pion, sigma, and rho,
but we also consider the $f_1$ meson as a means of determining the coupling
constant $G_f$.
The T-matrices for these four mesons can be written\footnote{
  The T-matrix for the pion can have additional structure
  generated by $\pi$-$a_1$ mixing,
  but we neglect this as a higher-order effect in this work.
}:
\begin{align}
  i{\mathcal{T}_\pi(q)}_{\alpha\beta,\gamma\delta}
  &=
  \frac{-2iG_\pi}{1+2G_\pi\Pi_{PP}(q^2)}
  (\gamma_5\tau_i)_{\alpha\beta}
  (\gamma_5\tau_i)_{\gamma\delta}
  \\
  i{\mathcal{T}_\sigma(q)}_{\alpha\beta,\gamma\delta}
  &=
  \frac{2iG_\pi}{1-2G_\pi\Pi_{SS}(q^2)}
  (1)_{\alpha\beta}
  (1)_{\gamma\delta}
  \\
  i{\mathcal{T}_\rho(q)}_{\alpha\beta,\gamma\delta}
  &=
  \frac{-2iG_\rho}{1+2G_\rho\Pi_{VV}(q^2)}
  \left[
    g^{\mu\nu} + 2G_\rho\Pi_{VV}(q^2)\frac{q^\mu q^\nu}{q^2}
    \right]
  (\gamma_\mu\tau_i)_{\alpha\beta}
  (\gamma_\nu\tau_i)_{\gamma\delta}
  \\
  i{\mathcal{T}_{f_1}(q)}_{\alpha\beta,\gamma\delta}
  &=
  \frac{-2iG_f}{1+2G_f\Pi_{AA}^{(T)}(q^2)}
  \left[
    g^{\mu\nu} + 2G_f\Pi_{AA}^{(T)}(q^2)\frac{q^\mu q^\nu}{q^2}
    \right]
  (\gamma_\mu\gamma_5)_{\alpha\beta}
  (\gamma_\nu\gamma_5)_{\gamma\delta}
  \,,
\end{align}
where the bubble diagrams are defined in App.~\ref{sec:bubbles}.
Since the mesons appear as poles in these T-matrices,
their masses are given by the conditions:
\begin{align}
  1 + 2G_\pi\Pi_{PP}(m_\pi^2) &= 0
  \label{eqn:pion:mass}
  \\
  1 - 2G_\pi\Pi_{SS}(m_\pi^2) &= 0
  \label{eqn:sigma:mass}
  \\
  1 + 2G_\rho\Pi_{VV}(m_\rho^2) &= 0
  \label{eqn:rho:mass}
  \\
  1 + 2G_f\Pi_{AA}^{(T)}(m_{f_1}^2) &= 0
  \label{eqn:f1:mass}
  \,.
\end{align}
The residues at the poles can be interpreted as effective
quark-meson coupling constants,
which allow us to find the properly normalized
homogeneous Bethe-Salpeter vertex functions:
\begin{align}
  \Gamma_\pi^i &= \sqrt{Z_\pi} \gamma_5 \tau_i \\
  \Gamma_\sigma^i &= \sqrt{Z_\sigma} \\
  \Gamma_\rho^i &= \sqrt{Z_\rho} \gamma^\mu \tau_i
  \,,
\end{align}
where
\begin{align}
  Z_\pi^{-1}
  =
  - \frac{\partial}{\partial q^2} \Pi_{PP}(q^2) \bigg|_{q^2=m_\pi^2}
  \\
  Z_\sigma^{-1}
  =
  + \frac{\partial}{\partial q^2} \Pi_{SS}(q^2) \bigg|_{q^2=m_\sigma^2}
  \\
  Z_\rho^{-1}
  =
  - \frac{\partial}{\partial q^2} \Pi_{VV}(q^2) \bigg|_{q^2=m_\rho^2}
  \,.
\end{align}

The NJL model is non-renormalizable,
owing to the four-fermi contact interaction.
To fully define the model,
it is necessary to introduce a regularization scheme.
We follow Refs. \cite{Ebert:1996vx,Hellstern:1997nv,Cloet:2014rja}
in using proper time regularization,
with both an infrared and an ultraviolet regulator.
The regularization proceeds formally through the substitution:
\begin{align}
  \frac{1}{X^n}
  =
  \frac{1}{(n-1)!} \int_0^\infty \mathrm{d}\tau\, \tau^{n-1} e^{-\tau X}
  \rightarrow
  \frac{1}{(n-1)!}
  \int_{1/\Lambda_{\mathrm{UV}}^2}^{1/\Lambda_{\mathrm{IR}}^2} \mathrm{d}\tau\,
  \tau^{n-1} e^{-\tau X}
  \,.
\end{align}
Only the UV regulator is necessary to make the model finite,
but the presence of an IR regulator ensures the two-body T-matrix
is always real, and prevents the decay of mesons into two quarks.
As is customary with non-remormalizable theories,
these regulators are kept finite as additional model parameters.

We adopt the NJL model parameters used in previous work \cite{Cloet:2014rja},
along with a value for $G_f$ which produces an $f_1$ pole in the T-matrix
with mass 1.28~GeV.
The model parameters we use are given in Tab.~\ref{tab:NJL:params}.

\begin{table}[t]
  \setlength{\tabcolsep}{0.5em}
  \renewcommand{\arraystretch}{1.3}
  \caption{
    NJL model parameters used in this work.
    All but $G_f$ are adopted from Ref. \cite{Cloet:2014rja}.
    $G_f$ is determined by requiring an $f_1$ pole at $m_{f_1}=1.28$~GeV
    in the NJL T-matrix.
    Couplings are in units GeV$^{-2}$,
    while the dressed quark mass $M$ and regulators are in units GeV.
  }
  \label{tab:NJL:params}
  \begin{tabular}{@{} cccccccc @{}}
    \toprule
    $ \Lambda_{\mathrm{IR}} $ &
    $ \Lambda_{\mathrm{UV}} $ &
    $ M $ &
    $ G_\pi=G_\eta $ &
    $ G_\rho $ &
    $ G_\omega $ &
    $ G_f $ &
    $ G_T $ \\
    \hline
    0.240 &
    0.645 &
    0.4 &
    19.0 &
    11.0 &
    10.4 &
    0.82 &
    0 \\
    \bottomrule
  \end{tabular}
\end{table}


\section{Gravitation and the EMT}
\label{sec:graviton}

The energy-momentum tensor (EMT) has long had strong ties to gravitation.
The equivalence between inertial and gravitational mass has long suggested
that energy is the charge upon which gravitation acts,
and this equivalence principle is currently canonized as one of the central
premises of general relativity
(and its extensions, such as Einstein-Cartan theory
\cite{Cartan:1923zea,Cartan:1924yea}).
Accordingly, it is helpful in theoretical investigations of the EMT---even
if gravitation is not the intended topic of study---to consider the physics
of graviton coupling.
If the EMT is the current upon which gravitons act,
then one can constrain the EMT of a system of interest using traditional
field theoretic methods that have been applied to other currents
such as the electromagnetic current.
Just as with coupling to a photon,
one can obtain a Ward-Takahashi identity for graviton coupling
and solve Dyson-Schwinger equations for the interaction between a graviton
and a fully dressed field in a strongly-interacting, non-perturbative regime.


\subsection{Gravitational Ward-Takahashi identity}

The gravitational Ward-Takahashi identity (GWTI) is
given by~\cite{Brout:1966oea}:
\begin{align}
  \Delta_\mu \Gamma^{\mu\nu}_G(p^\prime,p)
  =
  p^\nu S^{-1}(p^\prime) - p'^\nu S^{-1}(p)
  \label{eqn:WTI}
  \,.
\end{align}
This relation has applicability to fields of general spin if the
canonical EMT is used as the source of the gravitational field.
A proof of the general validity of Eq.~(\ref{eqn:WTI})
can be found in App.~\ref{sec:gwti}.

The canonical EMT for which Eq.~(\ref{eqn:WTI}) has general applicability
naturally arises as the Noether current
associated with spacetime translation symmetry.
This current is not symmetric in its indices for particles with intrinsic
spin~\cite{Leader:2013jra,Lorce:2018zpf,Florkowski:2018fap},
as a consequence of the generalized angular momentum
\begin{align}
  M^{\mu\alpha\beta}
  =
  x^\alpha T^{\mu\beta} - x^\beta T^{\mu\alpha}
  + S^{\mu\alpha\beta}
\end{align}
being a conserved quantity---in
particular, the Noether current associated with Lorentz transformations.
Here $S^{\mu\alpha\beta}$ is the {\sl intrinsic} generalized
angular momentum,
and as a consequence of the conservation laws
$\partial_\mu M^{\mu\alpha\beta} = 0$
and $\partial_\mu T^{\mu\nu} = 0$,
one has
$T^{\alpha\beta}-T^{\beta\alpha} = -\partial_\mu S^{\mu\alpha\beta}$.
For particles with spin, this is not expected to be zero,
so the EMT is not expected to be symmetric.

It is possible to obtain a symmetric, ``Belinfante-improved''
EMT by adding the divergence of a superpotential to the EMT
derived through Noether's theorem~\cite{Leader:2013jra}.
The resulting tensor is also a conserved current,
but there is no guarantee that it encodes exactly the same
physical properties as the Noether current associated with
spacetime translations.
It has been observed that adding a total derivative to the
EMT can in fact change the values of measurable quantities
such as the D-term~\cite{Hudson:2017xug}.
We thus choose not to add any total derivatives to the EMT
in this work.

One reason for preferring the Belinfante-improved EMT over the
canonical EMT is that general relativity assumes the EMT to be symmetric,
and spacetime to accordingly be torsion-free.
However, we do not {\sl a priori} know whether spacetime
is really torsion-free, and whether gravitons can couple
to the antisymmetric component of the EMT\footnote{
  Einstein-Cartan theory \cite{Cartan:1923zea} is a straightforward extension
  of general relativity that incorporates spacetime torsion and uses
  an asymmetric EMT as the source of gravitation.
}.
Since in this work we consider graviton coupling only as a theoretical
means of developing the formalism for calculating the EMT---which
is empirically accessed through other means,
such as DVCS \cite{Ji:1996nm}
and hadron pair production in diphoton collisions
\cite{Kumano:2017lhr}---we
find it most prudent to consider the graviton as capable of containing torsion
and proceed to use the canonical, asymmetric EMT.

As remarked above, one major consequence of using the canonical rather than
the Belinfante EMT is that  Eq. (\ref{eqn:WTI}) holds in general.
One can readily confirm this for simple examples of particles with spin,
such as an elementary free fermion, for which the gravitational vertex is:
\begin{align}
  \vcenter{\hbox{\includegraphics[scale=0.5]{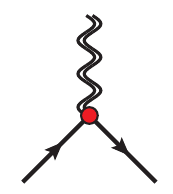}}}
  =
  \gamma_G^{\mu\nu}\left(k+\frac{\Delta}{2},k-\frac{\Delta}{2}\right)
  =
  \gamma^\mu k^\nu - g^{\mu\nu}(\slashed{k}-m)
  \label{eqn:graviton:elementary}
  \,.
\end{align}
On the other hand, the symmetrization of the vertex
$\gamma_G^{\{\mu\nu\}}$---which arises from using the Belinfante
EMT as the gravitating current in the spin-half case---does not
obey Eq.~(\ref{eqn:WTI}).
In fact, several authors \cite{DeWitt:1967uc,Bessler:1969py}
have found that the symmetrized graviton vertex obeys
a different Ward-Takahashi identity with a more complicated structure:
\begin{align}
  \Delta_\mu \Gamma^{\{\mu\nu\}}_G(p^\prime,p)
  =
  p^\nu S^{-1}(p^\prime) - p'^\nu S^{-1}(p)
  + \frac{1}{2i} \Delta_\mu \left[
    S^{-1}(p^\prime) \Sigma^{\mu\nu} - \Sigma^{\mu\nu} S^{-1}(p)
    \right]
  \label{eqn:WTI:DeWitt}
  \,,
\end{align}
where $\Sigma^{\mu\nu}$ is the generator of Lorentz transforms.
We consider the fact that it obeys a simpler WTI one of the
virtues of the canonical EMT.


\subsection{Graviton-quark vertices in NJL model}

Application of Noether's theorem to the full NJL model Lagrangian
of Eq. (\ref{eqn:lagrangian})
gives the following EMT:
\begin{align}
  T^{\mu\nu}(x)
  =
  \bar{\psi}(x) (i\gamma^\mu \overleftrightarrow{\partial}^\nu) \psi(x)
  -
  g^{\mu\nu}\bar{\psi}(x)(i\overleftrightarrow{\slashed{\partial}}-m)\psi(x)
  - \frac{1}{2} g^{\mu\nu}
  \sum_\Omega G_\Omega
  \big(\bar{\psi}(x)\Omega \psi(x)\big)
  \big(\bar{\psi}(x)\Omega \psi(x)\big)
  \label{eqn:EMT:NJL}
  \,,
\end{align}
which contains the non-linear contact interaction terms.
The presence of a four-fermi interaction term in the EMT entails the
existence of a five-point vertex function (4 quarks and 1 graviton line)
in addition to the usual three-point vertex (2 quarks and 1 graviton).

At the level of truncation we are considering here,
it is not necessary to dress the five-point vertex.
We do not dress the four-fermi contact interaction, after all,
and it would be inconsistent to dress the graviton's interaction
with the contact interaction while not dressing the interaction itself.

\begin{figure}
  \centering
  \includegraphics[width=0.5\textwidth]{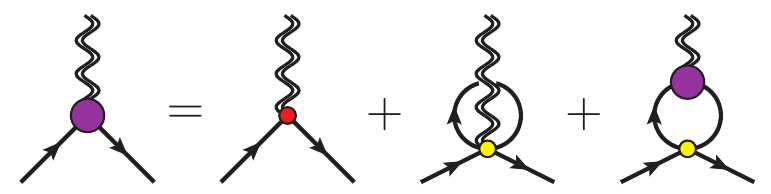}
  \caption{
    Inhomogeneous Bethe-Salpeter equation for the dressed quark-graviton
    vertex.
  }
  \label{fig:bse:Gqq}
\end{figure}

The effective Feynman rule for the five-point vertex can be read
off from the EMT in Eq.~(\ref{eqn:EMT:NJL}) directly:
\begin{align}
  \vcenter{\hbox{\includegraphics[scale=0.5]{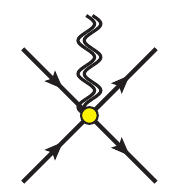}}}
  =
  \gamma_{Gqqqq}^{\mu\nu}
  =
  -g^{\mu\nu}
  \sum_{\Omega} 2G_\Omega \Omega\otimes\Omega
  \,.
\end{align}
We shall see presently that the existence of
this five-point vertex has non-trivial consequences,
including both a vacuum condensate contribution to the
Bethe-Salpeter equation for the three point vertex
(see second-from-the-right diagram of Fig.~\ref{fig:bse:Gqq}),
and the existence of a ``bicycle diagram''
in the calculation of bound state EMTs
(see rightmost diagram of Fig.~\ref{fig:graviton:pion}).
Crucially, these diagrams are both necessary for
energy-momentum conservation to be observed.

For the fully-dressed three-point vertex, we must solve an
inhomogeneous Bethe-Salpeter equation,
depicted in Fig.~\ref{fig:bse:Gqq}.
In contrast to the photon vertex BSE, there are two driving terms.
The second diagram from the right arises because it is possible
for the five-point vertex to contribute to the dressing of the
three-point vertex.
Most interestingly, this new driving term is directly proportional
to the vacuum condensate, which---as a consequence of the gap
equation, Eq.~(\ref{eqn:gap})---is proportional to the mass dressing $(M-m)$.
In particular, we have:
\begin{align}
  \vcenter{\hbox{\includegraphics[scale=0.5]{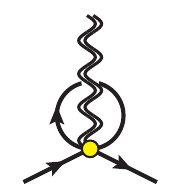}}}
  =
  2i G_\pi (2N_c)
  g^{\mu\nu}
  \int \frac{\mathrm{d}^4k}{(2\pi)^4}
  \mathrm{Tr}_D\left[ S(k) \right]
  =
  g^{\mu\nu} (M-m)
  \,,
\end{align}
meaning that the two driving terms together contribute
$\gamma^\mu k^\nu - g^{\mu\nu}(\slashed{k}-M)$
to the dressed graviton vertex.

When we take into consideration that the NJL model propagator
takes the same functional form as the bare propagator,
but with the current quark mass replaced by the dressed mass,
we observe the remarkable property that the driving terms in the
graviton vertex BSE already satisfy the gravitational WTI on their own.
This means the rightmost diagram in Fig.~\ref{fig:bse:Gqq} must
contribute a quantity transverse to the momentum transfer $\Delta$.
Additionally, since the NJL interaction kernel contains no
explicit momentum dependence and the $k$ dependence in the dressed
kernel is integrated over in the loop,
the rightmost diagram can only depend on $\Delta$.
These considerations,
and the fact that the EMT is $\mathsf{P}$ and $\mathsf{T}$ even,
gives us the following most general form for the dressed three-point vertex:
\begin{align}
  \vcenter{\hbox{\includegraphics[scale=0.5]{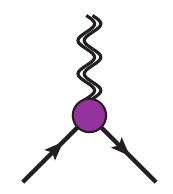}}}
  =
  \Gamma_{Gqq}^{\mu\nu}\left(k+\frac{\Delta}{2},k-\frac{\Delta}{2}\right)
  =
  \gamma^\mu k^\nu - g^{\mu\nu}(\slashed{k}-M)
  +
  \frac{\Delta^\mu\Delta^\nu - \Delta^2 g^{\mu\nu}}{4M} C_Q(t)
  +
  \frac{i\epsilon^{\mu\nu\Delta\sigma}\gamma_\sigma\gamma_5}{4} D^\prime_Q(t)
  \label{eqn:graviton:dressed}
  \,.
\end{align}
The functions $C_Q(t)$ and $D^\prime_Q(t)$ can be determined by algebraically
solving the BSE.
We find:
\begin{align}
  C_Q(t) &= \frac{ -8 G_\pi \Pi_{SG}(t) }{ 1 - 2G_\pi \Pi_{SS}(t) }
  \label{eqn:CQ}
  \\
  D^\prime_Q(t) &= \frac{ 2 G_f \Pi_{AA}^{(T)}(t) }{ 1 + 2G_f \Pi_{AA}^{(T)}(t) }
  \label{eqn:DQ}
  \,,
\end{align}
with the bubbles given in App.~\ref{sec:bubbles}.
Remarkably, $C_Q(t)$ has a pole in the timeline region when $t=m_\sigma^2$,
and $D_Q^\prime(t)$ has a timelike pole when $t=m_{f_1}^2$,
by comparison to the pole conditions
in Eqs.~(\ref{eqn:sigma:mass},\ref{eqn:f1:mass}).


\subsection{EMT decomposition and gravitational form factors}

Now that we have constructed the necessary formalism,
we proceed to consider matrix elements of the NJL model EMT
between momentum and spin eigenstates of a particular hadron, {\sl viz.},
$\langle p',\lambda'| T_{\mu\nu}(0) |p,\lambda\rangle$.
For conciseness, we refer to this matrix element as the EMT of the hadron.

\begin{figure}
  \centering
  \includegraphics[width=0.666\textwidth]{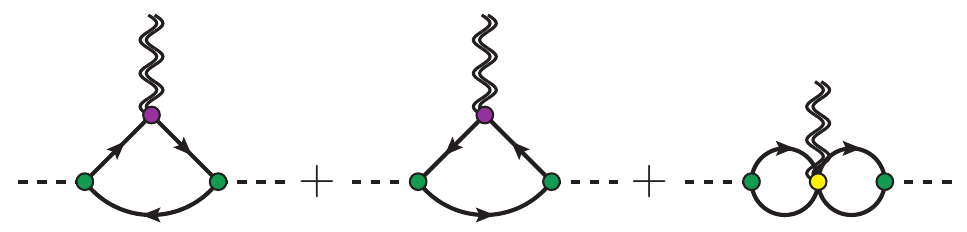}
  \caption{
    Diagrams contributing to meson-graviton coupling.
  }
  \label{fig:graviton:pion}
\end{figure}

To begin, we must consider three diagrams when calculating the EMT
of a meson in the NJL model.
These diagrams are given in Fig.~\ref{fig:graviton:pion}.
The first two (triangle) diagrams are completely analogous to the
diagrams contributing to the electromagnetic or axial vector current.
The third (bicycle) diagram is new to graviton interactions,
and is a consequence of the equivalence principle.
The four-fermi contact interaction is a sort of potential energy
in the NJL model, and gravitation couples to all energy in exactly
the same way.

The EMT of any hadron can be decomposed into a finite number of independent
algebraic structures,
each multiplied by a Lorentz-invariant function only of the invariant
momentum transfer $t=(p'-p)^2$.
These functions are called gravitational form factors (GFFs),
and are analogous to the electromagnetic form factors appearing in
decompositions of the electromagnetic current
$\langle p',\lambda'| j_\mu(0) |p,\lambda\rangle$.
Like with electromagnetic form factors,
the number of GFFs depends on the spin of the hadron,
and this number increases with increasing spin.

The most general form the spin-zero EMT can take
for a particular parton flavor $a$ is~\cite{Polyakov:2018zvc}:
\begin{align}
  \langle p^\prime \mid T^a_{\mu\nu}(0) \mid p \rangle
  &=
  2P_\mu P_\nu A_a(t)
  + \frac{1}{2}(\Delta_\mu \Delta_\nu - \Delta^2 g_{\mu\nu}) C_a(t)
  + 2 m_\pi^2 \bar{c}_a(t) g_{\mu\nu}
  \,,
\end{align}
where $a=q,g$,
$P = (p+p^\prime)/2$ is the average momentum between initial and final states,
and $\Delta = p^\prime-p$ is the momentum transfer.
The following sum rules follow from conservation of energy and momentum:
\begin{align}
  & \sum_{a=q,g} \bar{c}_{a}(t) = 0 \\
  & \sum_{a=q,g} A_{a}(0) = 1
  \,.
\end{align}
Since the NJL model does not contain gluons,
the former sum rule translates for us to $\bar{c}_q(t) = 0$.
Another rule, following from a low-energy pion
theorem~\cite{Novikov:1980fa,Voloshin:1980zf,Polyakov:2018zvc},
is:
\begin{align}
  \lim_{m_\pi\rightarrow0} \sum_{a=q,g} C_{a}(0) = -1
  \label{eqn:lowenergy}
  \,.
\end{align}
This is a rule that holds exactly only in the chiral limit,
and is a consequence of the pion being the Nambu-Goldstone boson
of chiral symmetry breaking.
Since the NJL model is a model of dynamical chiral symmetry breaking,
we should observe Eq.~(\ref{eqn:lowenergy}) when taking the pion mass to zero.
Moreover, one expects a quantity close to (although not exactly) $-1$ even
at the physical pion mass.

The most general form the spin-one EMT can take is
\cite{Cosyn:2019aio,Polyakov:2019lbq}:
\begin{align}
  \langle p^\prime, \lambda^\prime \mid T^a_{\mu\nu}(0) \mid p, \lambda \rangle
  &=
  -2P_\mu P_\nu\left[(\epsilon^{\prime*} \epsilon)
  \mathcal{G}^a_1(t)
  -
  \frac{(\Delta\epsilon^{\prime*} )(\Delta\epsilon )}{2m_\rho^2}
  \mathcal{G}^a_2(t)\right]
  \notag \\ &
  - \frac{1}{2}(\Delta_\mu \Delta_\nu - \Delta^2 g_{\mu\nu})
  \left[(\epsilon^{\prime*} \epsilon)
  \mathcal{G}^a_3(t)
  -
  \frac{(\Delta\epsilon^{\prime*} )(\Delta\epsilon )}{2m_\rho^2}
  \mathcal{G}^a_4(t)\right]
  +
  P_{\{\mu}\left( \epsilon^{\prime*}_{\nu\}} (\Delta \epsilon)
  - \epsilon_{\nu\}} (\Delta \epsilon^{\prime*}) \right)
  \mathcal{G}^a_5(t)
  \notag \\ &
  +
  \frac{1}{2} \left[
    \Delta_{\{\mu}\left( \epsilon^{\prime*}_{\nu\}} (\Delta\epsilon)
    + \epsilon_{\nu\}} (\Delta\epsilon^{\prime*}) \right)
    - \epsilon_{\{\mu}^{\prime*}\epsilon_{\nu\}} \Delta^2
    - g_{\mu\nu}(\Delta\epsilon^{\prime*})(\Delta\epsilon)
    \right]
  \mathcal{G}^a_6(t)
  \notag \\ &
  +\epsilon_{\{\mu}^{\prime*}\epsilon_{\nu\}} m_\rho^2  \mathcal{G}^a_7(t)
  + g_{\mu\nu} m_\rho^2 (\epsilon'^*\epsilon) \mathcal{G}^a_8(t)
  + \frac{1}{2}g_{\mu\nu}(\Delta\epsilon'^*)( \Delta\epsilon) \mathcal{G}^a_9(t)
   \notag \\ &
  +
  P_{[\mu}\left( \epsilon^{\prime*}_{\nu]} (\Delta \epsilon)
  - \epsilon_{\nu]} (\Delta \epsilon^{\prime*}) \right)
  \mathcal{G}^a_{10}(t)
  +
  \Delta_{[\mu}\left( \epsilon^{\prime*}_{\nu]} (\Delta \epsilon)
  + \epsilon_{\nu]} (\Delta \epsilon^{\prime*}) \right)
  \mathcal{G}^a_{11}(t)
  \,,
\end{align}
where $A^{\{\mu\nu\}} = \frac{1}{2}(A^{\mu\nu}+A^{\nu\mu})$
and $A^{[\mu\nu]} = \frac{1}{2}(A^{\mu\nu}-A^{\nu\mu})$.
The following sum rules follow from energy-momentum conservation,
and the assumption that the symmetric and antisymmetric components of the EMT
are separately conserved\footnote{
  This assumption is true of the gauge-invariant kinetic EMT
  used in \cite{Cosyn:2019aio},
  and is also of the canonical EMT of the quark field in the NJL model.
  However, the symmetric and antisymmetric components of the canonical EMT
  of the gluon field are not separately conserved in QCD.
  One should note that the gauge-invariant kinetic form of the gluon EMT
  does not give rise to a graviton vertex that satisfies the gravitational
  WTI of Eq.~(\ref{eqn:WTI}) that we employ in this work,
  while the canonical gluon EMT does.
}:
\begin{align}
  \sum_{a=q,g} \mathcal{G}_{7}^a(t) =
  \sum_{a=q,g} \mathcal{G}_{8}^a(t) =
  \sum_{a=q,g} \mathcal{G}_{9}^a(t) =
  \sum_{a=q,g} \mathcal{G}_{11}^a(t) = 0
  \,,
\end{align}
but since there are no gluons in the NJL model,
each of these form factors should be identically zero for the quarks.
We have two additional sum rules, the first also following from
energy-momentum conservation,
and the latter from angular momentum conservation:
\begin{align}
  & \sum_{a=q,g} \mathcal{G}_{1}^a(0) = 1 \\
  & \sum_{a=q,g} \mathcal{G}_{5}^a(0) = 2
  \,.
\end{align}
Lastly, there is a correspondence between
the GFF $\mathcal{G}_{10}(t)$ for a spin-one hadron
is related to its isoscalar axial form factors:
\begin{align}
  \mathcal{G}_{10}^q(t)
  &=
  -\widetilde{G}_1^q(t) + \frac{t}{m_\rho^2}\widetilde{G}_2^q(t)
  \label{eqn:rho:G10}
  \,.
\end{align}
This correspondence is a consequence of the QCD equation of motion
\cite{Leader:2013jra}:
\begin{align}
  i\overline{\psi} \gamma^{[\mu}\overleftrightarrow{\partial}^{\nu]}\psi
  =
  -
  \frac{1}{2}
  \epsilon^{\mu\nu\rho\sigma}
  \partial_\rho \Big(
    \overline{\psi} \gamma_\sigma \gamma_5 \psi
  \Big)
  \label{eqn:eom:5}
  \,.
\end{align}
However, Eq.~(\ref{eqn:eom:5}) also holds in the NJL model
(see App.~\ref{sec:aproof} for a proof),
and therefore Eq.~(\ref{eqn:rho:G10}) is also expected to hold
in the NJL model.


\subsection{Energy-momentum tensor of free quarks}

Although quarks are not asymptotic states in the NJL model
owing to the introduction of an infrared regulator,
we may ask what the EMT and gravitational form factors
of a free dressed quark would look like if we were to remove this regulator.
The most general from that the EMT of a spin-half particle can take
is~\cite{Lorce:2018egm}:
\begin{multline}
  \langle p^\prime,\lambda^\prime \mid T^a_{\mu\nu}(0) \mid p,\lambda \rangle
  =
  \bar{u}(p^\prime,\lambda^\prime)
  \bigg[
    \frac{P_\mu P_\nu}{M} A_a(t)
    + \frac{iP_{\{\mu}\sigma_{\nu\}\Delta}}{2M} [ A_a(t) + B_a(t) ]
    \\
    + \frac{\Delta_\mu\Delta_\nu-\Delta^2g_{\mu\nu}}{M}C_a(t)
    + Mg_{\mu\nu}\bar{c}_a(t)
    + \frac{iP_{[\mu}\sigma_{\nu]\Delta}}{2M} D_a(t)
    \bigg]
  u(p,\lambda)
  \,.
\end{multline}
Using the quark-graviton vertex in Eq.~(\ref{eqn:graviton:dressed})
and placing the quark on-shell gives,
once the quark flavors have been summed over,
$A_Q(t) = 1$, $B_Q(t) = 0$, $C_Q(t)$ as given in Eq.~(\ref{eqn:CQ}),
$\bar{c}_Q(t) = 0$, and $D_Q(t) = -1 + D_Q^\prime(t)$,
with $D_Q^\prime(t)$ given in Eq.~(\ref{eqn:DQ}).
This allows the functions $C_Q(t)$ and $-1+D_Q^\prime(t)$ we found by
solving the BSE to be interpreted as gravitational form factors of dressed quarks.

\begin{figure}
  \centering
  \includegraphics[width=0.6\textwidth]{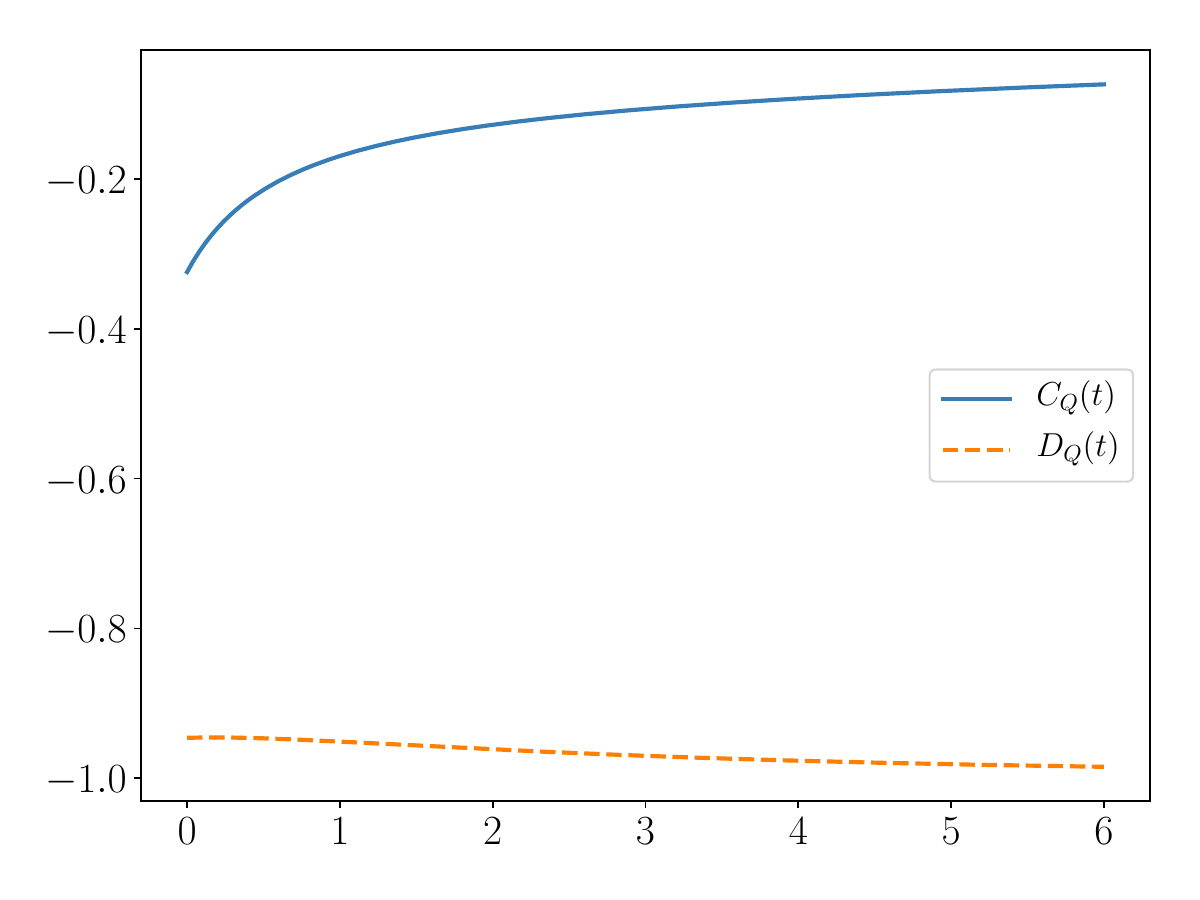}
  \caption{
    The gravitational form factors $C_Q(t)$ and $D_Q(t)$ of
    a dressed quark in the NJL model.
    $A_Q(t)=1$ and $B_Q(t)=0$ are not shown here,
    since they are trivial (undressed) in this model.
  }
  \label{fig:CQ}
\end{figure}

We have plotted in Fig.~\ref{fig:CQ} the two non-trivial GFFs of a dressed quark,
leaving out the trivial $A_Q(t)=1$ and $B_Q(t)=0$,
using the NJL model parameters given in Tab.~\ref{tab:NJL:params}.
In Fig.~\ref{fig:CQ}, both $C_Q(t)$ and $D_Q(t)$ can be seen to approach
their elementary values of $0$ and $-1$, respectively.
It should be noted that $C(t)=0$ for
an elementary free Dirac particle \cite{Hudson:2017oul},
and that a finite outcome for this form factor can only be produced
by interaction dynamics.


\section{Axial form factors}
\label{sec:axial}

\begin{figure}
  \centering
  \includegraphics[width=0.5\textwidth]{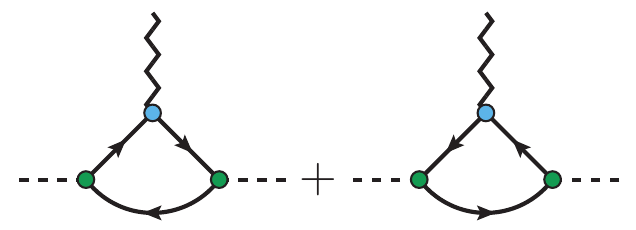}
  \caption{
    Diagrams contributing to the axial vector current of a meson.
  }
  \label{fig:axial:meson}
\end{figure}

In order to check Eq.~(\ref{eqn:rho:G10}) within the NJL model,
we must also calculate the axial form factors of the rho meson.
Since gravitation is an isoscalar interaction,
we look specifically at the isoscalar axial vector current.
The matrix element of the axial vector current takes the following
most general form for a spin-one particle \cite{Frederico:1992vm,Berger:2001zb}:
\begin{align}
  \langle p',\lambda' | J_5^{\mu}(0) | p,\lambda\rangle
  &=
  -2i \epsilon^{\mu \epsilon'^* \epsilon P} \widetilde{G}_1(t)
  -2i \epsilon^{\mu\Delta P\sigma}
  \frac{
    \epsilon_\sigma (\epsilon'^* \Delta)
    - \epsilon^{\prime*}_\sigma(\epsilon\Delta)
  }{ m_\rho^2 }
  \widetilde{G}_2(t)
  \,.
\end{align}
In order to calculate the matrix element in this equation,
we must evaluate the triangle diagrams in Fig.~\ref{fig:axial:meson}
with the axial vector vertex for a dressed quark,
and this vertex itself must be found through an inhomogeneous
Bethe-Salpeter equation.
In this section, we will find the axial form factors of the rho by first
finding the requisite axial vector vertex,
and in the process explore its properties and demonstrate that it satisfies
the axial Ward-Takahashi identity.

\subsection{Axial vector and pseudoscalar vertices}

\begin{figure}
  \centering
  \includegraphics[width=0.5\textwidth]{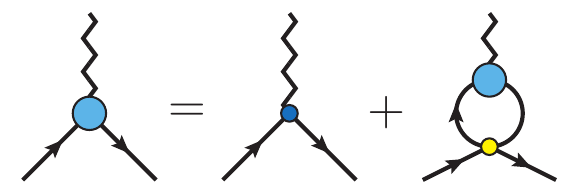}
  \caption{
    Diagrammatic representation of the inhomogeneous
    Bethe-Salpeter equation for either the pseudoscalar or
    axial vector vertex.
  }
  \label{fig:bse:axial}
\end{figure}

We begin by finding the dressed axial vector and pseudoscalar vertices
in the NJL model.
The relevant currents are isoscalar, and the driving terms for the
axial vector and pseudoscalar vertices are respectively
$\gamma^\mu\gamma_5$ and $\gamma_5$.
We denote the dressed vertices $\Gamma_5^{\mu}(p^\prime,p)$
for the axial vector and $\Gamma_5(p^\prime,p)$ for the pseudoscalar current.

The most general forms that the dressed vertices can take are:
\begin{align}
  \Gamma_5^{\mu}(p^\prime,p)
  &=
  a_1(t) \gamma^\mu\gamma_5
  + \frac{\Delta^\mu}{M}\gamma_5 a_2(t)
  + \frac{\Delta^\mu\slashed{\Delta}}{t}\gamma_5 a_3(t)
  \\
  \Gamma_5(p^\prime,p)
  &=
  g_1(t) \gamma_5 + \frac{\slashed{\Delta}}{M} g_2(t)
  \,,
\end{align}
where $a_i(t)$ and $g_i(t)$ are off-shell form factors
to be solved for algebraically.
The resulting solutions to the two BSEs are:
\begin{align}
  & g_1(t) =
  \frac{ 1 + 2G_f \Pi_{AA}^{(L)}(t) }{ \mathcal{D}_A(t) }
  \,,\qquad
  g_2(t) =
  \frac{ 2G_f \Pi_{PA}(t) }{ \mathcal{D}_A(t) }
  \\
  & a_1(t) =
  \frac{1}{ 1 + 2G_f\Pi_{AA}^{(T)}(t) }
  \,,\qquad
  a_2(t) =
  \frac{ -2G_\eta \Pi_{PA}(t) }{ \mathcal{D}_A(t) }
  \,,
  \\
  & a_3(t) =
  \frac{-2G_f}{1+2G_f\Pi_{AA}^{(L)}(t)} \left[
    \Big(\Pi_{AA}^{(L)}(t) - \Pi_{AA}^{(T)}(t)\Big) a_1(t)
    + \frac{t}{M^2} \Pi_{AP}(t) a_2(t)
    \right]
  \,,
\end{align}
where the bubbles are given in App.~\ref{sec:bubbles}, and
\begin{align}
  \mathcal{D}_A(t)
  &=
  \big(1+2G_\eta\Pi_{PP}(t)\big)\big(1+2G_f\Pi_{AA}^{(L)}(t)\big)
  - 4G_\eta G_f \frac{t}{M^2} \Pi_{AP}(t)\Pi_{PA}(t)
  \,.
\end{align}


\subsection{Axial vector Ward-Takahashi identity}

The axial vector and pseudoscalar vertices we have found
should satisfy an axial vector Ward-Takahashi identity,
which can be stated as \cite{Adler:1969gk}:
\begin{align}
  \Delta_\mu \Gamma_5^{\mu}(p^\prime,p)
  =
  S^{-1}(p^\prime)\gamma_5 + \gamma_5 S^{-1}(p)
  +
  2m\Gamma_5(p^\prime,p)
  \label{eqn:WTI:axial}
  \,.
\end{align}
Here, $m$ is the current quark mass rather than the dressed quark mass.
One can immediately observe that:
\begin{align}
  a_1(t) + a_3(t)
  &=
  \frac{1+2G_\eta\Pi_{PP}(t)}{\mathcal{D}_A(t)}
  \\
  \Delta_\mu \Gamma_5^{\mu}(p^\prime,p)
  &=
  \slashed{\Delta}\gamma_5\Big(a_1(t)+a_3(t)\Big)
  + \frac{t}{M} \gamma_5 a_2(t)
  =
  \left[
    \frac{1+2G_\eta\Pi_{PP}(t)}{\mathcal{D}_A(t)}
    \slashed{\Delta}
    -
    \frac{t}{M}
    \frac{ 2G_\eta \Pi_{PA}(t) }{ \mathcal{D}_A(t) }
    \right]
  \gamma_5
  \\
  S^{-1}(p^\prime)\gamma_5 + \gamma_5 S^{-1}(p)
  &=
  \slashed{\Delta}\gamma_5 - 2M\gamma_5
  \\
  2m\Gamma_5(p^\prime,p)
  &=
  2m \left[
    \frac{ 1 + 2G_f \Pi_{AA}^{(L)}(t) }{ \mathcal{D}_A(t) }
    +
    \frac{ 2G_f \Pi_{PA}(t) }{ \mathcal{D}_A(t) }
    \frac{\slashed{\Delta}}{M}
    \right]
    \gamma_5
  \,,
\end{align}
meaning the axial vector WTI requires that:
\begin{align}
  \frac{m}{M}
  =
  \frac{1+2G_\eta\Pi_{PP}(t)-\mathcal{D}_A(t)}{2G_f\Pi_{PA}(t)}
  =
  \frac{2\mathcal{D}_A(t)-2G\eta\frac{t}{M^2}\Pi_{PA}(t)}{1+2G_f\Pi_{AA}^{(L)}(t)}
  \label{eqn:require}
  \,.
\end{align}
This can be proved, but requires a little work to show.
The equality between the second and third expressions in Eq.~(\ref{eqn:require})
can be shown using a little algebra using $\Pi_{PA}(t)=-\Pi_{AP}(t)$
and $\Pi_{AA}^{(L)}(t) = -2\Pi_{PA}(t)$.
The equality between the first and second expressions requires:
\begin{align}
  \frac{m}{M}
  &=
  \frac{
    -2G_f\Pi_{AA}^{(L)}(t)\Big(1+2G_\eta\Pi_{PP}(t)\Big)
    +
    4G_\eta G_f \frac{t}{M^2}\Pi_{PA}(t)\Pi_{AP}(t)
  }{ 2 G_f \Pi_{PA}(t) }
  \notag \\ &=
  1 + 2G_\eta\Pi_{PP}(t) + G_\eta \frac{t}{M^2}\Pi_{AP}(t)
\end{align}
The remaining bubbles can be found to evaluate to:
\begin{align}
  \Pi_{PP}(t)
  &=
  4i(2N_c)
  \int_0^1 \mathrm{d}x\,
  \int \frac{\mathrm{d}^4k}{(2\pi)^4}
  \left[
    - \frac{1}{k^2 - M^2 + x(1-x)t + i0}
    + \frac{2x(1-x)t}{[k^2 - M^2 + x(1-x)t + i0]^2}
    \right]
  \\
  \Pi_{AP}(t)
  &=
  - 4i(2N_c)
  \int_0^1 \mathrm{d}x\,
  \int \frac{\mathrm{d}^4k}{(2\pi)^4}
  \frac{M^2}{[k^2 - M^2 + x(1-x)t + i0]^2}
  \,,
\end{align}
where we leave the regularization scheme unspecified and implicit,
except to assume that several basic operations are allowed, namely:
(1) differentiation and integration with respect to variables
other than $k$ commutes with the $k$ integral, and
(2) one can cancel factors of $k^2-B(M,x,t)$ between the numerator
and denominator and still obtain the same result.
These restrictions, it should be noted, are met by the proper time
regularization scheme we employ.
With these rules, we can find that:
\begin{align}
  \frac{t}{M^2} \Pi_{AP}(t) + 2\Pi_{PP}(t)
  &=
  - 4i (2N_c)
  \int \frac{\mathrm{d}^4k}{(2\pi)^4}
  \int_0^1 \mathrm{d}x\,
  \left[
    \frac{2}{k^2 - M^2 + x(1-x)t + i0}
    + \frac{1-4x(1-x)t}{[k^2 - M^2 + x(1-x)t + i0]^2}
    \right]
  \notag \\ &=
  - 8i (2N_c)
  \int \frac{\mathrm{d}^4k}{(2\pi)^4}
  \int_0^1 \mathrm{d}x\,
  \left[
    \frac{1}{k^2 - M^2 + x(1-x)t + i0}
    - \frac{x(1-2x)t}{[k^2 - M^2 + x(1-x)t + i0]^2}
    \right]
  \notag \\ &=
  - 8i (2N_c)
  \int \frac{\mathrm{d}^4k}{(2\pi)^4}
  \int_0^1 \mathrm{d}x\,
  \frac{\partial}{\partial x}
  \left[
    \frac{x}{k^2 - M^2 + x(1-x)t + i0}
    \right]
  \notag \\ &=
  - 8i (2N_c)
  \int \frac{\mathrm{d}^4k}{(2\pi)^4}
  \frac{1}{k^2 - M^2 + i0}
  \,.
\end{align}
Comparison with the gap equation, Eq.~(\ref{eqn:gap}), gives us:
\begin{align}
  \frac{t}{M^2} \Pi_{AP}(t) + 2\Pi_{PP}(t)
  =
  -\left(\frac{M-m}{G_\pi M}\right)
  \,,
\end{align}
and therefore the axial vector WTI requires that:
\begin{align}
  \frac{m}{M}
  =
  1 - \frac{G_\eta}{G_\pi}\frac{M-m}{M}
  \,,
\end{align}
which holds true if and only if $G_\eta = G_\pi$.
This requirement is not surprising, since $\mathrm{U}(1)_A$ invariance
is the symmetry responsible for the partial conservation of isoscalar
axial symmetry, and therefore for the isoscalar axial vector WTI.
It was remarked in \cite{Cloet:2014rja} that $G_\eta = G_\pi$ is
a necessary condition for the NJL model Lagrangian (\ref{eqn:lagrangian})
to satisfy $\mathrm{U}(1)_A$ symmetry,
and the necessity of this equality in the final step of our proof
is simply a manifestation of that fact.


\section{Results}
\label{sec:results}

We now present results for the pion and rho meson EMTs as calculated in
the two-flavor NJL model.
Since the EMT is an isoscalar quantity,
and since we take $m_u=m_d$ in this work,
the up and down quark contributions to the EMT are equal,
and we present the meson EMT obtained after summing over both quark flavors.


\subsection{Pion and sigma}

\begin{figure}
  \includegraphics[width=0.6\textwidth]{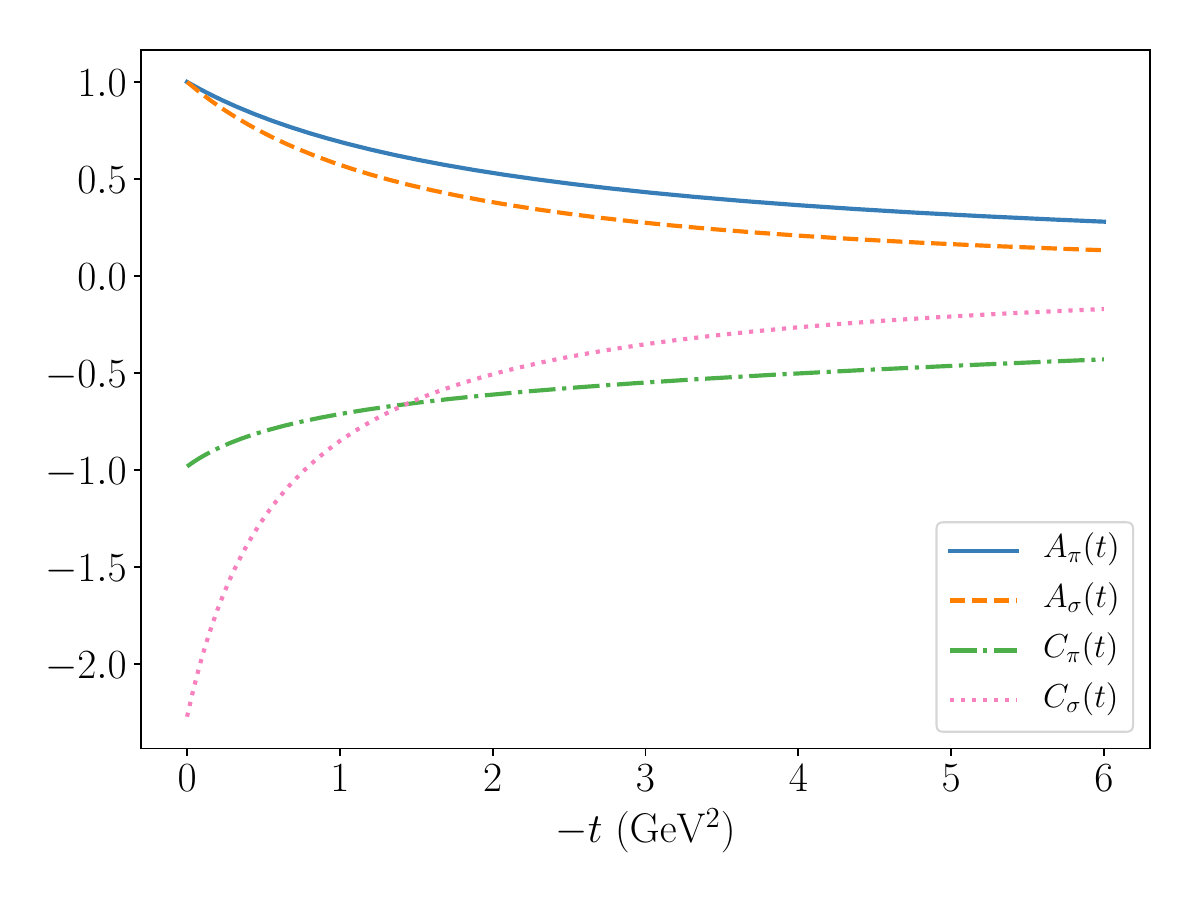}
  \caption{
    Gravitational form factors of the pion and sigma meson.
  }
  \label{fig:pion:plot}
\end{figure}

The GFFs of the pion and sigma meson
have been plotted in Fig.~\ref{fig:pion:plot}.
One can see immediately that the expected constraints appear to hold.
In particular, we have $A_{\pi,\sigma}(0)=1$ and  $C_\pi(0)\approx -1$.
Not included in the plot are the results $\bar{c}_{\pi,\sigma}(t)=0$,
which are found to hold exactly.
In addition, we find that in the chiral limit, $C_\pi(0)=-1$ exactly,
satisfying the low-energy pion theorem of Eq.~(\ref{eqn:lowenergy}).

It's worth stressing the importance of the ``quark D term''
[$C_Q(t)$ in Eq.~(\ref{eqn:graviton:dressed}), not $D_Q(t)$]
in satisfying the low-energy pion theorem.
This has implications for the calculation of
generalized parton distributions (GPDs).
The leading-twist GPDs $H_\pi^{q,g}(x,\xi,t)$
are related to GFFs through the
second Mellin moment\footnote{
  This is true for the gluon GPD is the Ji convention \cite{Ji:1998pc}
  is used. If the Diehl convention \cite{Diehl:2003ny} is used,
  the gluon GFFs come from half the first Mellin moment of the GPD.
}, {\sl viz.},
\begin{align}
  \int_{-1}^1 \mathrm{d}x\,
  H_\pi^a(x,\xi,t)
  &=
  A_\pi^a(t) + \xi^2 C_\pi^a(t)
  \,,
\end{align}
which is a specific case of the more general property that
Mellin moments of GPDs produce matrix elements of local operators.

If a bare non-local operator is used to calculate the GPDs,
then its moments can produce only matrix elements of bare local operators.
For the second moment in particular, this will give the ``$++$''
component of the bare three-point graviton vertex.
As a consequence, if one does not dress the non-local operator
defining the GPDs, an incorrect value will be inferred for $C_a(t)$.

To illustrate this, we define ``partially dressed'' gravitational
form factors as those that incorporate the vacuum condensate
diagram, but for which $C_Q(t)$ is set to zero.
We note that we need to include the vacuum condensate diagram
for consistency, as otherwise the gravitational WTI is violated.
Additionally, this diagram has no bearing on the Mellin moments
of leading-twist GPDs, since its contribution contains a factor
$g^{\mu\nu}$, and $g^{++}=0$.
On the other hand, $C_Q(t)$ should appear in the Mellin moment
with a factor $\xi^2$, so we expect $C_\pi(t)$ to be incorrect
with this term missing.

\begin{figure}
  \includegraphics[width=0.6\textwidth]{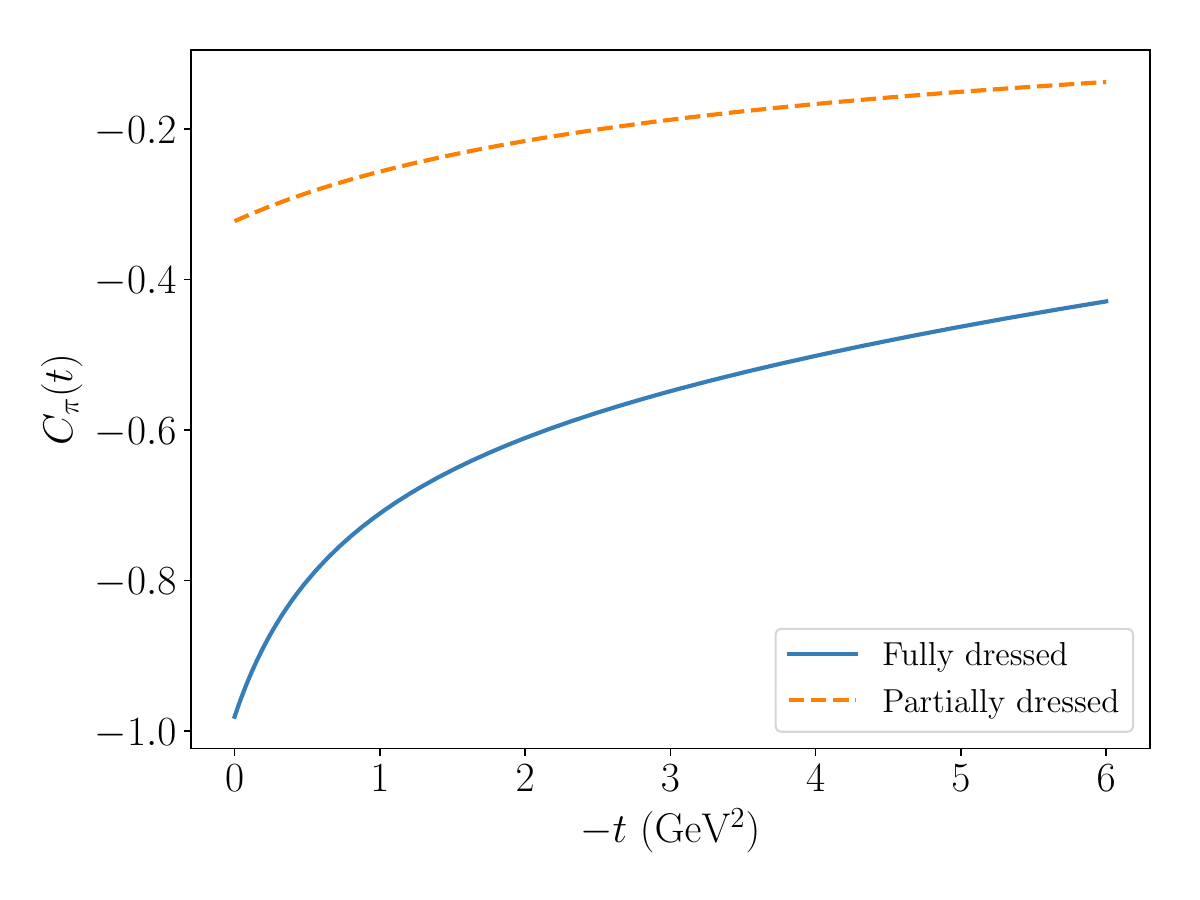}
  \caption{
    The ``D-term'' form factor $C_\pi(t)$ for the pion.
    (Solid curve) Full calculation.
    (Dashed curve) With the quark D-term $C_Q(t)$ neglected.
  }
  \label{fig:pion:plot:C}
\end{figure}

In Fig.~\ref{fig:pion:plot:C}, we show the difference between
$C_\pi(t)$ as calculated in the NJL model
with and without the quark D-term $C_Q(t)$ folded in.
The contrast is actually stark, with about two-thirds of the
pion D-term coming from the quark dressing.
We repeat and reemphasize that the ``partially dressed''
curve in this plot is the $C_\pi(t)$ that one would find from
taking the second Mellin moment of the leading-twist pion GPD
if it were calculated using a bare non-local operator,
and that this speaks to the necessity of dressing operators
in GPD calculations.

We illustrate this point further by finding analytic expressions for
$C_Q(0)$ and $C_\pi(0)$ within proper time regularization.
We find:
\begin{align}
  C_\pi(t=0)
  &=
  \frac{1}{3}
  \frac{
    \int_0^1\mathrm{d}x\, \int\mathrm{d}\tau \left\{
      -[1-6C_Q(0)]
      \frac{1}{\tau}
      +
      [1 + 6C_Q(0)]
      \frac{x(1-x)m_\pi^2}{M^2-x(1-x)m_\pi^2}
      \right\}
    e^{-(M^2-x(1-x)m_\pi^2)\tau}
  }{
    \int_0^1\mathrm{d}x\, \int\mathrm{d}\tau \left\{
      \frac{1}{\tau}
      +
      \frac{x(1-x)m_\pi^2}{M^2-x(1-x)m_\pi^2}
      \right\}
    e^{-(M^2-x(1-x)m_\pi^2)\tau}
  }
  \xrightarrow[m_\pi\rightarrow0]{}
  -\frac{1}{3} + 2C_Q(0)
  \\
  C_Q(t=0)
  &=
  -\frac{1}{3}
  \frac{
    G_\pi \frac{M^2}{\pi^2}
    \int_0^1\mathrm{d}x\, \int\mathrm{d}\tau \frac{1}{\tau}
  }{
    \frac{m}{M}
    +
    G_\pi \frac{M^2}{\pi^2}
    \int_0^1\mathrm{d}x\, \int\mathrm{d}\tau \frac{1}{\tau}
  }
  \xrightarrow[m\rightarrow0]{}
  -\frac{1}{3}
  \,,
\end{align}
where the $\tau$ integration is from
$\Lambda_{\mathrm{UV}}^{-2}$ to $\Lambda_{\mathrm{IR}}^{-2}$.
One can indeed see that $C_\pi(0)=-1$ exactly in the chiral limit.
Additionally, we instead get exactly $-\frac{1}{3}$ if the contribution
from $C_Q(0)$ is neglected---meaning a full two-thirds of the expected
value come from the quark dressing.

The necessity for fulling dressing the relevant operator to observe
the low-energy pion theorem---either
the three-point graviton vertex or the bilocal light cone operator---draws
an analogy to other results of dynamical chiral symmetry breaking,
such as the dressing of quark mass.
These non-perturbative phenomena both require a self-consistent solution
of the relevant Dyson-Schwinger equations to manifest,
and both result in significant changes to physical, observable
quantities that are not constrained by conservation laws.
This can be contrasted with electric charge,
which remains unaltered between the bare electromagnetic vertex
and the dressed vertex found by solving the inhomogeneous BSE.

In contrast to the pion, the sigma meson is not a Nambu-Goldstone boson
and its ``D-term'' is not constrained in any {\sl a priori} fashion.
One can observe that $C_\sigma(0)=-2.27\not\approx-1$.

On another point,
it is worth mentioning that the vanishing of $\bar{c}_{\pi,\sigma}(t)$
is possible only when all three diagrams have been included.
Each of the diagrams on its own
makes a non-zero contribution to the spin-zero meson EMT.
The triangle diagrams contribute
$\frac{1}{2} Z_\pi \Pi_{PP}(m_\pi^2)$
or
$\frac{1}{2} Z_\sigma \Pi_{SS}(m_\sigma^2)$
each to $\bar{c}_\pi(t)$ or $\bar{c}_\sigma(t)$, respectively
(with direct application of the WTI of Eq.~(\ref{eqn:WTI})
being the most straightforward way to find this result),
while the bicycle diagram contributes
$2 G_\pi Z_\pi \Big( \Pi_{PP}(m_\pi^2) \Big)^2$
or
$-2 G_\sigma Z_\sigma \Big( \Pi_{SS}(m_\sigma^2) \Big)^2$
to the same form factor.
The three diagrams sum to
\begin{align}
  2m_\pi^2 \bar{c}_\pi(t)
  =
  Z_\pi \Pi_{PP}(m_\pi^2) \left[
    1 + 2G_\pi \Pi_{PP}(m_\pi^2)
    \right]
  = 0
  \\
  2m_\sigma^2
  \bar{c}_\sigma(t)
  =
  Z_\sigma \Pi_{SS}(m_\pi^2) \left[
    1 - 2G_\sigma \Pi_{SS}(m_\sigma^2)
    \right]
  = 0
  \,,
\end{align}
which both vanish because of the pion and sigma
mass-shell conditions in Eqs.~(\ref{eqn:pion:mass},\ref{eqn:sigma:mass}).

\subsubsection{Spin-zero meson mass radius}

With the EMT of the spin-zero mesons in hand, it is also possible to study
their static mechanical properties.
It is conventional in much of the
literature \cite{Polyakov:2002yz,Lorce:2017wkb}
to define a Breit frame EMT through:
\begin{align}
  \langle T^a_{\mu\nu}\rangle_{\mathrm{Breit}}(\mathbf{r})
  \equiv
  \int \frac{\mathrm{d}^3\boldsymbol{\Delta}}{2P^0(2\pi)^3}
  e^{-i(\boldsymbol{\Delta}\mathbf{r})}
  \langle p^\prime \,|\, T^a_{\mu\nu}(0) \,|\, p \rangle
  \bigg|_{\mathbf{P}=0}
  \label{eqn:EMT:Breit}
  \,,
\end{align}
where the requirement that $\mathbf{P} = \mathbf{p}^\prime-\mathbf{p}=0$
gives us $P^0 = \sqrt{m_\pi^2 + \boldsymbol{\Delta}^2/4}$
and $\Delta^0 = 0$.
One can then evaluate moments of
$\langle T_{\mu\nu}\rangle_{\mathrm{Breit}}(\mathbf{r})$
to obtain static properties such as the mass radius.

Interpretation of Eq.~(\ref{eqn:EMT:Breit}) as an actual
spatial distribution has been called into question \cite{Miller:2018ybm}.
Despite this, one could still formally define a mass radius in terms
of the Breit frame ``density,'' as is often done with the proton charge radius.
Nonetheless, we run into problems for the pion.
The limit $\boldsymbol{\Delta}\rightarrow0$
must be taken when evaluating multipole moments,
which requires the existence of a rest frame for the particle.
Thus, Eq.~(\ref{eqn:EMT:Breit}) cannot be used in the chiral limit.

One may alternatively define a two-dimensional spatial distribution
using light cone quantization \cite{Dirac:1949cp},
where the spatial dimensions are the transverse light cone coordinates.
The transverse spatial distribution is then:
\begin{align}
  \langle T^a_{\mu\nu}\rangle_{\mathrm{LC}}(\mathbf{r}_\perp)
  \equiv
  \int \frac{\mathrm{d}^2\boldsymbol{\Delta}_\perp}{2P^+(2\pi)^2}
  e^{-i(\boldsymbol{\Delta}_\perp\mathbf{r}_\perp)}
  \langle p^\prime \,|\, T^a_{\mu\nu}(0) \,|\, p \rangle
  \bigg|_{\Delta^+=0}
  \label{eqn:EMT:LC}
  \,,
\end{align}
where the requirement $\Delta^+=0$ implies $P^+ = p^+ = p'^+$.
Since in light cone quantization $P^+$ is kinematic and
$P^-$ is dynamical \cite{Dirac:1949cp,Brodsky:1997de}\footnote{
  The roles of $P^+$ and $P^-$ in \cite{Dirac:1949cp}
  are the opposite as in this work.
},
$P^+$ has no dependence on $\boldsymbol{\Delta}$,
and thus no dependence on $t$, by contrast to $P^0$ in the Breit frame.
We shall see the significance of this presently.

The mass radius can be found as the mean value of either
$\mathbf{r}^2$ weighted by
$\frac{1}{2P^0}\langle T^{00}\rangle_{\mathrm{Breit}}(\mathbf{r})$,
or $\mathbf{r}_\perp^2$ weighted by
$\frac{1}{2P^+}\langle T^{++}\rangle_{\mathrm{LC}}(\mathbf{r}_\perp)$.
We find:
\begin{align}
  \langle r^2 \rangle_{\mathrm{Breit}}
  &=
  \sum_{a=q,g}
  \lim_{\boldsymbol{\Delta}\rightarrow0}
  -\frac{1}{P^0}
  \nabla_\Delta^2 \left[
    \frac{1}{2P^0}
    \langle p^\prime \,|\, T_a^{00}(0) \,|\, p \rangle
    \bigg|_{\Delta^0=0}
    \right]
  =
  6 \frac{\mathrm{d}A_{\pi,\sigma}(t)}{\mathrm{d}t} \bigg|_{t=0}
  - \frac{3}{4m_{\pi,\sigma}^2}\left[ A_{\pi,\sigma}(0) + 2C_{\pi,\sigma}(0) \right]
  \label{eqn:pion:radius:Breit}
  \\
  \langle r_\perp^2 \rangle_{\mathrm{LC}}
  &=
  \sum_{a=q,g}
  \lim_{\boldsymbol{\Delta}\rightarrow0}
  -\frac{1}{P^+}
  \nabla_{\Delta_\perp}^2 \left[
    \frac{1}{2P^+}
    \langle p^\prime \,|\, T_a^{++}(0) \,|\, p \rangle
    \bigg|_{\Delta^+=0}
    \right]
  =
  4 \frac{\mathrm{d}A_{\pi,\sigma}(t)}{\mathrm{d}t} \bigg|_{t=0}
  \label{eqn:pion:radius:LC}
  \,.
\end{align}
These radii differing (beyond a factor of the number of spatial dimensions)
ultimately amounts to the mass form factor
being different in the Breit frame and on the light cone,
respectively $P^0 A(t)$ and $P^+ A(t)$.
In the relevant frames, $P^0$ has $t$ dependence while $P^+$ does not.
This is analogous to the electric charge distribution of
the nucleon \cite{Miller:2007uy} or deuteron \cite{Carlson:2008zc}
differing between the Breit frame and the light cone.
In the case of the nucleon, the Sachs form factor $G_E(t)$ describes
the electric charge distribution in the Breit frame,
while the Dirac form factor $F_1(t)$ does instead on the light cone.

A remarkable property of the Breit frame radius (\ref{eqn:pion:radius:Breit})
is that it remains finite even for point particles,
for which $\frac{\mathrm{d}A(t)}{\mathrm{d}t}=0$.
By contrast, the light cone radius (\ref{eqn:pion:radius:LC}) is zero
for point particles.
The former can be understood as owing to spatial distributions not being
invariant under Lorentz boosts,
while the latter is due to the fact that transverse boosts in
light cone coordinates are Galilean \cite{Brodsky:1997de}.
In more detail, spatial distributions in the Breit frame are found by
integrating over a collection of reference frames where the system of interest
is in motion by different amounts,
but without applying any corrections to counteract Lorentz contractions.
These relativistic corrections are intrinsically accounted for by using light
cone coordinates, however.

\begin{table}[t]
  \setlength{\tabcolsep}{0.5em}
  \renewcommand{\arraystretch}{1.3}
  \caption{
    Mean squared mass radius of spin-zero mesons in the NJL model,
    both using the Breit frame and light cone prescriptions.
    For the pion, an empirical value for the light cone mass radius
    extracted from KEKB data is included for comparison.
    All values are in fm.
  }
  \label{tab:pion:radius}
  \begin{tabular}{@{}ccccc@{}}
    \toprule
    ~ &
    Breit frame &
    Light cone &
    Empirical light cone \cite{Kumano:2017lhr} [see text]
    \\
    \hline
    Pion &
    1.28 &
    0.27 &
    $0.26\sim0.32$ \\
    Sigma &
    0.56 &
    0.32 &
    ~ \\
    \bottomrule
  \end{tabular}
\end{table}

Because of the $\mathcal{O}(m_\pi^{-2})$ term
in Eq.~(\ref{eqn:pion:radius:Breit}), the Breit
frame radius of the pion blows up in the chiral limit.
On the other hand, Eq.~(\ref{eqn:pion:radius:LC}) remains finite
at zero pion mass.
Both equations can be used at physical pion mass,
but produce staggeringly different values for the mass radius.
Numerical values can be found in Tab.~\ref{tab:pion:radius}.
For the empirical value of the pion mass radius,
we look to the extraction in \cite{Kumano:2017lhr},
where a dispersive analysis of KEKB data for
$\gamma^*\gamma\rightarrow\pi^0\pi^0$
was done to extract gravitational form factors.
The formula used in \cite{Kumano:2017lhr} was the same as our
Eq.~(\ref{eqn:pion:radius:LC}) for the squared light cone pion mass radius,
but with a factor 6 instead of 4.
We have thus scaled down the range of $0.32\sim0.39$~fm
reported in \cite{Kumano:2017lhr} by a factor $\sqrt{2/3}$.
The NJL model result for this radius agrees with the empirical range,
but falls on the low end.

In Ref.~\cite{Cloet:2014rja}, the Breit frame charge radius of the
pion was found to be 0.62~fm, which is significantly smaller than
the Breit frame mass radius we have found.
However, as we have discussed, the Breit frame mass radius made artificially
large by $\mathcal{O}(m_\pi^{-1})$ terms
that are not present in the light cone mass radius.
The pion charge radius of Ref.~\cite{Cloet:2014rja}, when scaled by
$\sqrt{2/3}$, gives a light cone charge radius of 0.51~fm,
which is instead larger than the light cone mass radius.
The Breit frame and light cone prescriptions for radii thus suggest
strongly divergent pictures of the relative distribution of mass and charge
in the pion.
Since doubt has been cast on the interpretation of the Breit frame
radius as the moment of an actual density \cite{Miller:2018ybm},
the picture painted by light cone coordinates seems more plausible.

Because the form factor $A_\sigma(t)$ falls off faster than $A_\pi(t)$
(see Fig.~\ref{fig:pion:plot}),
the light cone mass radius of the sigma is larger than that of the pion.
On the other hand, because of its greater mass,
the sigma meson has a smaller Breit frame radius than the pion.
There is a stark difference between the results in these two frames,
again demonstrating the magnitude and importance of correctly accounting
for relativistic effects---by, for instance, using light cone coordinates
in defining spatial distributions.

Finally, we remark again that getting the correct value of $C_\pi(0)$
by fully dressing the quark-graviton vertex is vital here.
If one neglects the ``quark D-term,''
or equivalently obtains the GFFs through Mellin moments of bare GPDs,
so that $C_\pi(0)\approx-\frac{1}{3}$, then one finds
$\langle r^2 \rangle_{\mathrm{Breit}} < 0$
at the physical pion mass---an obvious absurdity
that violates the weak energy condition \cite{Hawking:1973uf}.


\subsection{Rho meson}

The rho meson, as a spin-one hadron, has many more GFFs than the pion.
There are 11 GFFs total,
with 4 of these being non-conserved \cite{Cosyn:2019aio}.
As required from the lack of gluons in the NJL model,
the four non-conserved GFFs vanish:
$\mathcal{G}_7=\mathcal{G}_8=\mathcal{G}_9=\mathcal{G}_{11}=0$.
There are thus seven non-zero GFFs to consider.

Three of the rho GFFs have direct analogues to the pion GFFs,
in particular,
$\mathcal{G}_1(t)\sim A_\pi(t)$,
$\mathcal{G}_3(t)\sim C_\pi(t)$, and
$\mathcal{G}_8(t)\sim -2\bar{c}_\pi(t)$.
The behavior of these form factors is remarkably similar to those
of the pion.
Firstly, we find $\mathcal{G}_1(0)=1$, as is required by momentum conservation.
Additionally, we curiously find $\mathcal{G}_3(0)\approx-1$,
and even more curiously find that this becomes exactly $-1$ in the chiral limit.
Since the rho meson is not a Nambu-Goldstone boson,
we do not know of any theorems requiring that this be the case,
as we did for $C_\pi(0)$.
Lastly, as with $\bar{c}_\pi(t)$ for the pion,
we find that including the contribution of the bicycle diagram
(rightmost diagram in Fig.~\ref{fig:graviton:pion})
is necessary for $\mathcal{G}_8(t)$ to fully vanish.

Among the new conserved form factors, $\mathcal{G}_5(t)$ describes the
spatial distribution of total angular momentum.
An angular momentum sum rule \cite{Abidin:2008ku}
requires that $\mathcal{G}_5(0)=2$,
and we satisfy this sum rule in the NJL model.
The non-conserved form factor $\mathcal{G}_7(t)$ also contributes to
this spatial distribution, but vanishes in the NJL model.
The remaining form factors contribute only to higher multipole moments
of the energy-momentum tensor.

\begin{figure}
  \includegraphics[width=0.6\textwidth]{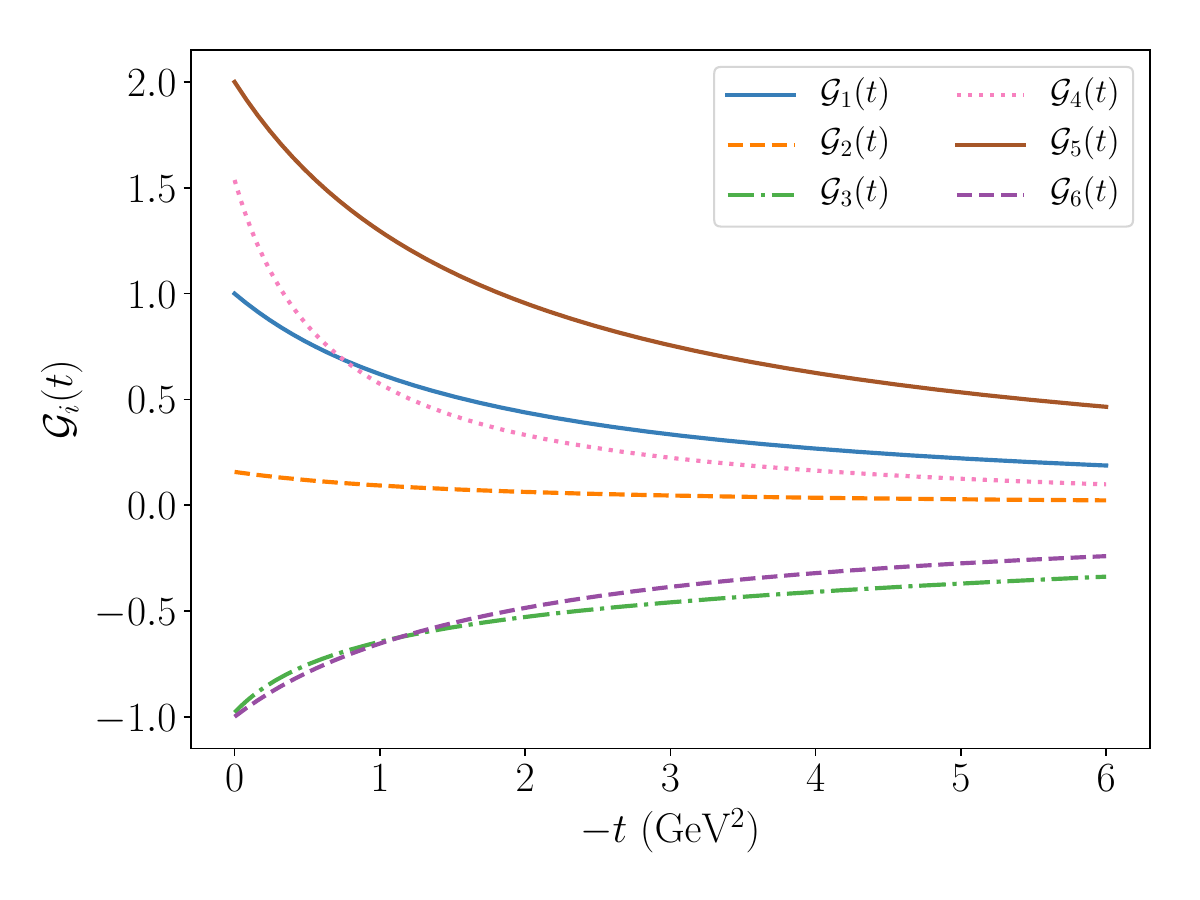}
  \caption{
    The six non-zero gravitational form factors
    of the rho meson appearing in the symmetric component
    of its energy-momentum tensor.
  }
  \label{fig:rho:plot}
\end{figure}

Of the seven non-zero GFFs, the six $\mathcal{G}_{1-6}(t)$ appear in
the symmetric component of the energy-momentum tensor.
These six form factors are of special phenomenological interest,
since they can be found from second Mellin moments of leading-twist
generalized parton distributions \cite{Cosyn:2018thq,Cosyn:2019aio},
which can be measured in hard exclusive reactions such as DVCS and DVMP.
These six GFFs have been plotted in Fig.~\ref{fig:rho:plot}.

\begin{table}[t]
  \setlength{\tabcolsep}{0.5em}
  \renewcommand{\arraystretch}{1.3}
  \caption{
    Static properties of the rho meson in the NJL model.
    Electromagnetic properties taken from the BSE calculation
    (without pion cloud effects) for $\rho^+$ from \cite{Cloet:2014rja}.
    All radii are in fm,
    the mass quadrupole moment is in units of $m_\rho$-fm$^2$,
    and the electric quadrupole moment is in $e$-fm$^2$.
  }
  \label{tab:rho:radius}
  \begin{tabular}{@{}ccccc@{}}
    \toprule
    ~ &
    $ \sqrt{\langle r^2 \rangle_{\mathrm{mass}}} $ &
    $ \sqrt{\langle r^2 \rangle_{\mathrm{elec.}}} $ &
    $ \mathcal{Q}_{\mathrm{mass}} $ &
    $ \mathcal{Q}_{\mathrm{elec.}} $ \\
    \hline
    Breit frame &
    0.45 &
    0.67 &
    -0.0224 &
    -0.0200 \\
    Light cone &
    0.25 &
    0.45 &
    ~ & ~ \\
    \bottomrule
  \end{tabular}
\end{table}

In \cite{Cosyn:2019aio},
the Breit frame multipole moments of the spin-one EMT were found.
The mean squared mass radius and gravitational quadrupole moment are:
\begin{align}
  \langle r^2 \rangle_{\mathrm{mass}}
  &=
  6 \frac{\mathrm{d}\mathcal{G}_1(t)}{\mathrm{d}t}\bigg|_{t=0}
  + \frac{1}{m_\rho^2} \left[
    -\frac{7}{4}\mathcal{G}_1(0)
    -\mathcal{G}_2(0)
    -\frac{3}{2}\mathcal{G}_3(0)
    +\mathcal{G}_5(0)
    +\frac{1}{2}\mathcal{G}_6(0)
    \right]
  \\
  \mathcal{Q}_{\mathrm{mass}}
  &=
  \frac{1}{m_\rho} \left[
    -\mathcal{G}_1(0)
    -\mathcal{G}_2(0)
    +\mathcal{G}_5(0)
    +\frac{1}{2}\mathcal{G}_6(0)
    \right]
  \,,
\end{align}
where we have neglected the non-conserved form factors,
since they are zero in the NJL model.
Since the Breit frame density cannot be literally interpreted as an actual
spatial density \cite{Miller:2018ybm},
we also determine the light cone transverse mass radius as a point of contrast,
which we find to be:
\begin{align}
  \langle r_\perp^2 \rangle_{\mathrm{LC}}
  &=
  4 \frac{\mathrm{d}\mathcal{G}_1(t)}{\mathrm{d}t}\bigg|_{t=0}
  + \frac{1}{m_\rho^2} \left[
    \frac{2}{3}\mathcal{G}_1(0)
    -\frac{2}{3}\mathcal{G}_2(0)
    -\frac{2}{3}\mathcal{G}_5(0)
    -\frac{1}{3}\mathcal{G}_6(0)
    \right]
  \,.
\end{align}
Since the light cone density is a two-dimensional quantity,
we will defer exploration of the quadrupole moment---both quantitative
and conceptual---to a future work on the light cone interpretation
of the EMT.

The numerical values have been computed and tabulated
in Tab.~\ref{tab:rho:radius},
along with the equivalent electric (Coulomb) quantities \cite{Cloet:2014rja}.
The quadrupole moments are remarkably close when comparable units are used,
but the electric charge radius is larger than the mass radius,
for both the Breit frame and light cone radii.
This suggests a highly inhomogeneous distribution of electric charge
in the rho meson,
as could occur (for instance) in a configuration with a small
negatively-charged core surrounded by a shell of positive charge.
Curiously, this occurs even for a positive rho meson---consisting of
positively charged up and anti-down quarks---but we remark that the
dressed quarks themselves have spatially extended electric charge distributions.

\begin{figure}
  \includegraphics[width=0.6\textwidth]{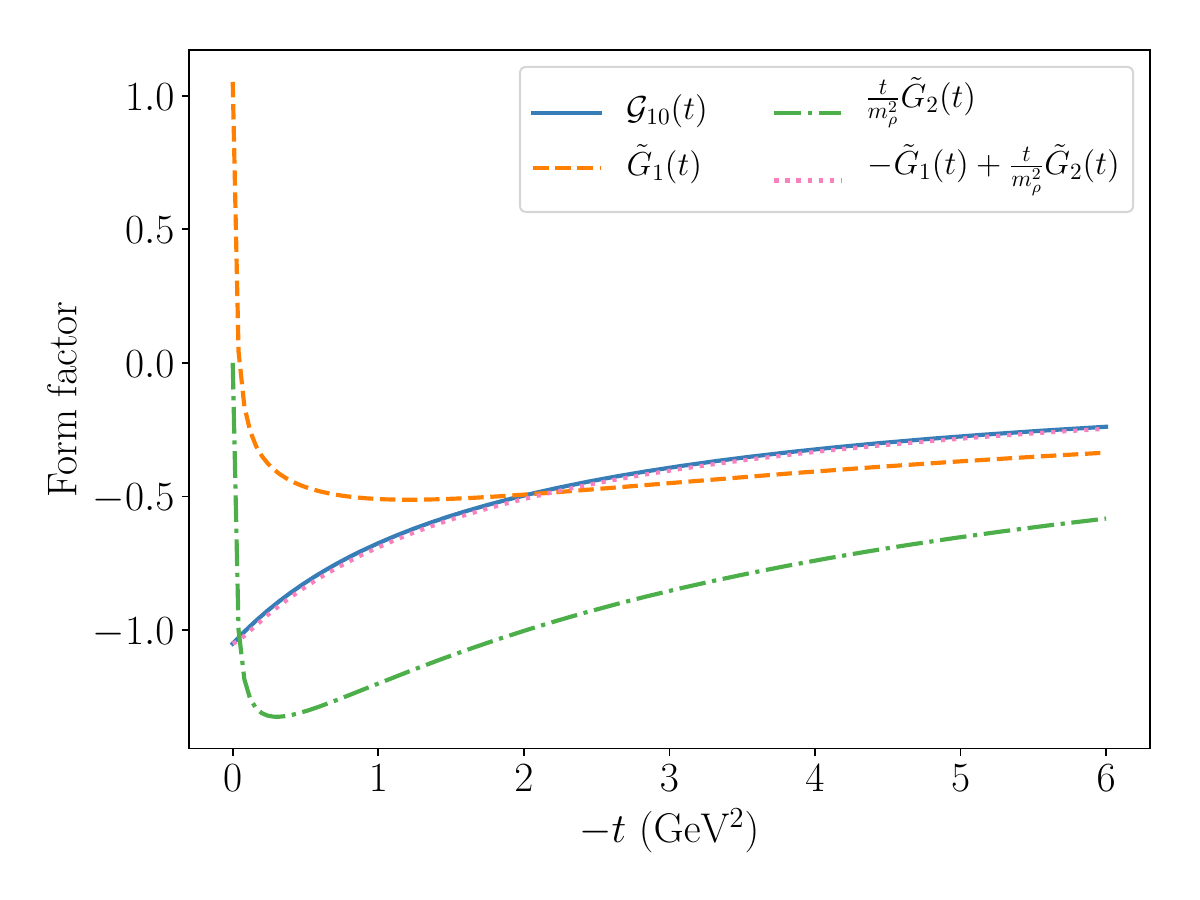}
  \caption{
    The gravitational form factor $\mathcal{G}_{10}(t)$ of the rho meson,
    compared with the rho axial form factors.
    The form factor $\tilde{G}_2(t)$ has been weighted by
    $t/m_\rho^2$ to make it comparable to the other form factors
    on the same plot.
  }
  \label{fig:rho:plot:G10}
\end{figure}

The conserved, non-zero GFF $\mathcal{G}_{10}(t)$
appears in the antisymmetric component of the EMT.
In both QCD and the NJL model,
this GFF has a special relationship with the axial form factors,
given in Eq.~(\ref{eqn:rho:G10}).
To demonstrate this correspondence,
we plot the relevant form factors in Fig.~\ref{fig:rho:plot:G10}.
Because of this relation,
$\mathcal{G}_{10}(t)$ encodes information
about the distribution of quark spin,
with the quantity $-\frac{1}{2}\mathcal{G}_{10}(0)=0.526$
giving the proportion of the rho meson's total angular momentum
carried by quark spin.
If we take $G_f=0$, we instead get $0.556$,
a number consistent with previous work
on the rho meson in the NJL model~\cite{Ninomiya:2017ggn}
that implicitly assumed $G_f=0$.


\section{Conclusions}
\label{sec:concl}

In this work, we have found the gravitational form factors
appearing in the decompositions of the canonical energy-momentum tensor
for the pion, sigma, and tho mesons in the NJL model.
In the process of obtaining these, we proved a gravitational
Ward-Takahashi identity for the canonical EMT of fields with arbitrary spin,
which takes a simpler form than the equivalent WTI for the Belinfante EMT.
This simpler WTI makes consistency cross-checks between the dressed
graviton vertex and dressed propagator within model calculations easier,
and will also extend to models with spin-one constituents.

We found that the non-linear four-fermi interaction of the NJL model
entails a five-point gravitational vertex
(with four quark lines and one graviton line)
in addition to the usual three-point vertex (with two quark lines and one graviton).
This finding is a result of the equivalence principle,
and is necessary for conservation of energy and momentum to be observed.
In solving the Bethe-Salpeter equation for the dressed three-point vertex,
we found it was necessary to include contributions from the five-point vertex
for the gravitational WTI to be satisfied.
Additionally, the five-point vertex contributed to meson EMT calculations
directly in the form of a bicycle diagram,
which was needed in order for the non-conserved GFFs
$\bar{c}(t)$ and $\mathcal{G}_8(t)$ to vanish.

The necessity of the five-point vertex suggests difficulties in the
prospect of calculating gravitational form factors in QCD.
The QCD Lagrangian contains quark-gluon, three-gluon, four-gluon,
and (in a covariant gauge) ghost-gluon vertices in its Lagrangian,
and the equivalence principle---encoded by the presence of
$-g^{\mu\nu}\mathcal{L}$ in the EMT---requires that the graviton
be able to couple directly to every one of these vertices.
Inclusion of all these graviton interactions is likely necessary for
energy-momentum conservation to be observed, just as inclusion
of the five-point vertex was in the NJL model.
Moreover, consistent solution of Dyson-Schwinger equations for the dressings
of these vertices is likely necessary for the gravitational WTI to be satisfied.

The meson GFF results demonstrate several fascinating properties of
not just the mesons themselves, but the field theoretical framework
which gives rise to them.
A low-energy pion theorem requires $C_\pi(0)\approx-1$,
with the approximation becoming exact in the chiral limit.
The NJL model correctly reproduces this
(including the exactness in the chiral limit),
and we found fully dressing the three-point graviton vertex
by solving its inhomogeneous Bethe-Salpeter equation to be necessary
to reproduce this behavior---an observation that has implications for
model calculations of generalized parton distributions of the pion.

We also observed the importance of correctly accounting for relativistic effects
when describing spatial properties of light mesons,
especially the pion---and more especially when considering the chiral limit.
The Breit frame mass radius of the pion is significantly larger than its
transverse light cone radius, only the latter of which is even finite
in the chiral limit.
We concur with previous literature
that light cone coordinates are necessary to meaningfully define spatial distributions.
We find that the light cone mass for the pion predicted by the NJL model
agrees with a phenomenological extraction from KEKB data.

In future work, we plan to extend the methods developed here to baryons.
The results found here will be directly applicable within a quark-diquark model,
where scalar and axial vector diquarks mimic closely the structure of pions
and rho mesons.
Additionally, the gravitational WTI we derived can be used as a
consistency cross-check for the off-shell diquark-graviton vertex.

\section*{Acknowledgements}

We would like to thank
Rafael Badui, Wim Cosyn, Sabrina Cotogno, and Cédric Lorcé
for illuminating discussions that helped contribute to our investigation.
This work was supported by the U.S.\ Department of Energy, Office of Science,
Office of Nuclear Physics, contract no.\ DE-AC02-06CH11357.
AF was supported by an LDRD initiative at Argonne National Laboratory
under Project No.\ 2017-058-N0.


\appendix


\section{Bubbles in the NJL model}
\label{sec:bubbles}

\begin{figure}
  \centering
  \includegraphics[width=0.5\textwidth]{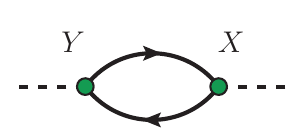}
  \caption{
    Bubble diagram.
  }
  \label{fig:bubble}
\end{figure}

The bubbles are defined using the following convention:
\begin{align}
  \Pi_{XY}(s)
  =
  i (2N_c) \int\frac{\mathrm{d}^4k}{(2\pi)^4} \mathrm{Tr}_D \left[
    \Omega_X S(k) \Omega_Y S(k-p)
    \right]
  \label{eqn:bubble}
\end{align}
where $\Omega_X$ and $\Omega_Y$ represent particular vertices
that appear in the bubble diagrams.
A diagrammatic depiction of the bubble is given in Fig.~\ref{fig:bubble},
where the factor $i^2$ from the propagators cancels the factor $(-1)$ appearing
because the diagram has a closed fermion loop.
An overall factor $i$ is included in the definition (\ref{eqn:bubble})
since this factor typically appears in diagrammatic equations involving
bubbles as sub-diagrams,
and this factor additionally makes the bubbles purely real
below the two-particle production threshold.
Specific bubbles that appear in this work are defined as follows:
\begin{align}
  \Pi_{PP}(q^2)
  &=
  (2N_c) i
  \int\frac{\mathrm{d}^4k}{(2\pi)^4}
  \mathrm{Tr}_D \left[
    \gamma_5 S(k) \gamma_5 S(k-q)
    \right]
  \label{eqn:bubble:PP}
  \\
  \Pi_{SS}(p^2)
  &=
  (2 N_c) i
  \int\frac{\mathrm{d}^4k}{(2\pi)^4}
  \mathrm{Tr}\left[
    S(k) S(k-p)
    \right]
  \label{eqn:bubble:SS}
  \\
  \left( g^{\mu\nu} - \frac{q^\mu q^\nu}{q^2} \right)
  \Pi_{VV}(q^2)
  &=
  (2N_c) i
  \int\frac{\mathrm{d}^4k}{(2\pi)^4}
  \mathrm{Tr}_D \left[
    \gamma^\mu S(k) \gamma^\nu S(k-q)
    \right]
  \label{eqn:bubble:VV}
  \\
  \left( g^{\mu\nu} - \frac{q^\mu q^\nu}{q^2} \right)
  \Pi_{AA}^{(T)}(q^2)
  + \frac{q^\mu q^\nu}{q^2} \Pi_{AA}^{(L)}(q^2)
  &=
  (2N_c) i
  \int\frac{\mathrm{d}^4k}{(2\pi)^4}
  \mathrm{Tr}_D \left[
    \gamma^\mu\gamma_5 S(k) \gamma^\nu\gamma_5 S(k-q)
    \right]
  \label{eqn:bubble:AA}
  \\
  \left( \frac{p^2 g^{\mu\nu} - p^\mu p^\nu}{M} \right)
  \Pi_{SG}(p^2)
  &=
  (2N_c) i
  \int \frac{\mathrm{d}^4k}{(2\pi)^4}
  \mathrm{Tr}\left[
    S(k) \gamma_{Gqq}^{\mu\nu}(k,k-p) S(k-p)
    \right]
  \label{eqn:bubble:SG}
  \,.
\end{align}

\section{Proof of the gravitational Ward-Takahashi identity}
\label{sec:gwti}

We here prove the gravitational Ward-Takahashi identity
given in Eq.~(\ref{eqn:WTI}),
first proved for spin-zero fields in \cite{Brout:1966oea},
is true for fields with arbitrary spin,
provided the graviton couples to the canonical energy-momentum tensor.

We must start with a definition of the fully dressed
3-point gravitational vertex function $\Gamma^{\mu\nu}(p^\prime,p)$.
The definition we use is:
\begin{align}
  \int \mathrm{d}^4x\, \mathrm{d}^4y\, \mathrm{d}^4z\,
  e^{-i[(py)-(p^\prime z)+(qx)]}
  \langle0| \mathtt{T}\{ T^{\mu\nu}(x) \phi_{r}(y) \phi_{s}(z) \} |0\rangle
  \notag \\ \equiv
  -i(2\pi)^4 \delta^4(p-p^\prime+q)
  \big( iS_{ss^\prime}(p^\prime) \big)
  \Gamma^{\mu\nu}_{s^\prime r^\prime}(p^\prime,p)
  \big( iS_{r^\prime r}(p) \big)
  \label{eqn:vertex:define}
  \,,
\end{align}
where $p$ and $p^\prime$ are respectively the initial and final momentum
of a quantum of a field $\phi$,
$S$ is the fully-dressed propagator of said field,
$T^{\mu\nu}$ is the {\sl canonical} energy-momentum tensor (EMT),
and we explicitly notate the internal degrees of freedom of $\phi$ with
indices $r,s$.

We begin by contracting the left-hand side of Eq.~(\ref{eqn:vertex:define})
with $q_\mu$. Let us use $L^{\mu\nu}_{rs}(p^\prime,p,q)$ as shorthand for
the entire LHS, for better use of space. We have:
\begin{align}
  q_\mu L^{\mu\nu}_{rs}(p^\prime,p,q)
  &=
  \int \mathrm{d}^4x\, \mathrm{d}^4y\, \mathrm{d}^4z\,
  q_\mu
  e^{-i[(py)-(p^\prime z)+(qx)]}
  \langle0| \mathtt{T}\{ T^{\mu\nu}(x) \phi_{r}(y) \phi_{s}(z) \} |0\rangle
  \notag \\ &=
  i
  \int \mathrm{d}^4x\, \mathrm{d}^4y\, \mathrm{d}^4z\,
  \Big(
  \partial_\mu^{(x)}
  e^{-i[(py)-(p^\prime z)+(qx)]}
  \Big)
  \langle0| \mathtt{T}\{ T^{\mu\nu}(x) \phi_{r}(y) \phi_{s}(z) \} |0\rangle
  \notag \\ &=
  -i
  \int \mathrm{d}^4x\, \mathrm{d}^4y\, \mathrm{d}^4z\,
  e^{-i[(py)-(p^\prime z)+(qx)]}
  \Big(
  \partial_\mu^{(x)}
  \langle0| \mathtt{T}\{ T^{\mu\nu}(x) \phi_{r}(y) \phi_{s}(z) \} |0\rangle
  \Big)
  \,.
\end{align}
Now, in differentiating this time-ordered product, we note that
$\partial_\mu T^{\mu\nu}(0) = 0$ as a consequence of Noether's theorem.
(We note that differentiation with respect to the {\sl first} index is
crucial here, as conservation with respect to the second index is
not guaranteed.)
Thus, any $x$ dependence in the time-ordered product can only come from
the time-ordering itself.
We have, explicitly:
\begin{align}
  \partial_\mu^{(x)}
  \langle0| \mathtt{T}\{ T^{\mu\nu}(x) \phi_r(y) \phi_s(z) \} |0\rangle
  &=
  \langle0| \mathtt{T}\left\{
    [T^{0\nu}(x), \phi_r(y)] \delta(x_0-y_0) \phi_s(z)
    + [T^{0\nu}(x), \phi_s(z)] \delta(x_0-z_0) \phi_r(y)
    \right\} |0\rangle
  \notag \\ &=
  \langle0| \mathtt{T}\left\{
    i\partial^\nu_{(y)} \phi_r(y) \phi_s(z) \delta^4(x-y)
    + i\partial^\nu_{(z)} \phi_s(z) \phi_r(y) \delta^4(x-z)
    \right\} |0\rangle
  \,,
\end{align}
where we have used the canonical commutation relation
\begin{align}
  [T^{0\mu}(x),\phi_r(y)]\delta(x_0-y_0) = i\partial^\mu \phi_r(y) \delta^4(x-y)
  \label{eqn:commute}
  \,,
\end{align}
from which $P^\mu = \int\mathrm{d}^3x\, T^{0\mu}(x)$ follows.
(Note that such a commutation relation does not hold for $T^{\mu0}(x)$,
giving us a second point at which privileging the first index is important.)
From the definition of the propagator, we obtain:
\begin{align}
  \partial_\mu^{(x)}
  \langle0| \mathtt{T}\{ T^{\mu\nu}(x) \phi_r(y) \phi_s(z) \} |0\rangle
  =
  i \partial^\nu_{(y)} S_{rs}(y-z) \delta^4(x-y)
  + i \partial^\nu_{(z)} S_{rs}(y-z) \delta^4(x-z)
  \,.
\end{align}
This enables us to find:
\begin{align}
  q_\mu L^{\mu\nu}_{rs}(p^\prime,p,q)
  &=
  \int \mathrm{d}^4x\, \mathrm{d}^4y\, \mathrm{d}^4z\,
  e^{-i[(py)-(p^\prime z)+(qx)]}
  \Big(
  \partial^\nu_{(y)} S_{rs}(y-z) \delta^4(x-y)
  + \partial^\nu_{(z)} S_{rs}(y-z) \delta^4(x-z)
  \Big)
  \notag \\ &=
  -
  \int \mathrm{d}^4x\, \mathrm{d}^4y\, \mathrm{d}^4z\,
  e^{-i[(py)-(p^\prime z)+(qx)]}
  \Big(
  \partial^\nu_{(z)} S_{rs}(y-z) \delta^4(x-y)
  + \partial^\nu_{(y)} S_{rs}(y-z) \delta^4(x-z)
  \Big)
  \notag \\ &=
  -
  \int \mathrm{d}^4x\, \mathrm{d}^4y\, \mathrm{d}^4z\,
  e^{-i[(py)-(p^\prime z)+(qx)]}
  \Big(
  -i p^{\prime\nu}
  S_{rs}(y-z) \delta^4(x-y)
  +
  i p^\nu
  S_{rs}(y-z) \delta^4(x-z)
  \Big)
  \,.
\end{align}
At this point, we compare the RHS of Eq.~(\ref{eqn:vertex:define})
and integrate over $q$. We have:
\begin{align}
  \int \frac{\mathrm{d}^4q}{(2\pi)^4}
  q_\mu L^{\mu\nu}_{rs}(p^\prime,p,q)
  &=
  -i
  \big( iS_{ss^\prime}(p^\prime) \big)
  \Gamma^{\mu\nu}_{s^\prime r}(p^\prime,p)
  \big( iS_{r^\prime r}(p) \big)
  \notag \\ &=
  i
  \int \mathrm{d}^4x\, \mathrm{d}^4y\, \mathrm{d}^4z\,
  e^{-i[(py)-(p^\prime z)]}
  \Big(
  p^{\prime\nu}
  S_{rs}(y-z) \delta^4(x-y) \delta^4(x)
  -
  p^\nu
  S_{rs}(y-z) \delta^4(x-z) \delta^4(x)
  \Big)
  \notag \\ &=
  i
  \int \mathrm{d}^4y\,
  \Big(
  p^{\prime\nu}
  S_{rs}(-y)
  e^{i(p^\prime y)}
  -
  p^\nu
  S_{rs}(y)
  e^{-i(py)}
  \Big)
  \notag \\ &=
  i\Big(
  p^{\prime\nu}
  S_{rs}(p^\prime)
  -
  p^\nu
  S_{rs}(p)
  \Big)
  \,.
\end{align}
Now, by cancelling out the factors $i$, multiplying by the inverse propagators,
and dropping the $r$ and $s$ indices, we get Eq.~(\ref{eqn:WTI}), as required.


\section{Proof of axial-gravitational correspondence in NJL model}
\label{sec:aproof}

To prove Eq.~(\ref{eqn:eom:5}) holds in the NJL model,
we follow the derivation in Sec.~3.2 of \cite{Leader:2013jra} for QCD.
Using the NJL model Lagrangian (\ref{eqn:lagrangian}),
we obtain the following equations of motion:
\begin{align}
  i\overrightarrow{\slashed{\partial}} \psi
  &=
  \left[
    m + \sum_\Omega G_\Omega (\overline\psi\Omega\psi) \Omega
    \right]
    \psi
  \equiv \hat{M} \psi
  \\
  i \overline\psi \overleftarrow{\slashed{\partial}}
  &=
  - \overline\psi \left[
    m + \sum_\Omega G_\Omega (\overline\psi\Omega\psi) \Omega
    \right]
  = \overline\psi \hat{M}
  \,.
\end{align}
From here, we use Eqs.~(129) of \cite{Leader:2013jra} with $A_\mu=0$,
which can be stated:
\begin{align}
  \sigma^{\mu\nu}\overrightarrow{\slashed{\partial}}
  &=
  2 \gamma^{[\nu} \overrightarrow{\partial}^{\mu]}
  + i\epsilon^{\mu\nu\rho\sigma} \gamma_\sigma\gamma_5 \overrightarrow{\partial}_\rho
  \notag \\
  \overleftarrow{\slashed{\partial}} \sigma^{\mu\nu}
  &=
  2 \overleftarrow{\partial}^{[\nu} \gamma^{\mu]}
  + i\epsilon^{\mu\nu\rho\sigma} \gamma_\sigma\gamma_5 \overleftarrow{\partial}_\rho
  \,.
\end{align}
With this, we find that:
\begin{align}
  2 \overline\psi \gamma^{[\nu}i\overleftrightarrow{\partial}^{\mu]} \psi
  =
  \overline\psi [\hat{M}, \sigma^{\mu\nu}] \psi
  +
  \epsilon^{\mu\nu\rho\sigma} \partial_\rho
  \Big( \overline\psi \gamma_\sigma \gamma_5 \psi \Big)
  \,.
\end{align}
What remains to be shown is that
$\overline\psi [\hat{M}, \sigma^{\mu\nu}] \psi = 0$.
Using the identities:
\begin{align}
  [1, \sigma^{\mu\nu}] &= 0 \\
  [\gamma_5, \sigma^{\mu\nu}] &= 0 \\
  [\gamma^\pi, \sigma^{\mu\nu}]
  &=
  2i \Big( g^{\mu\pi} \gamma^\nu - g^{\nu\pi} \gamma^\mu \Big)
  \\
  [\gamma^\pi\gamma_5, \sigma^{\mu\nu}]
  &=
  2i \Big( g^{\mu\pi} \gamma^\nu - g^{\nu\pi} \gamma^\mu \Big) \gamma_5
  \\
  [\sigma^{\pi\rho},\sigma^{\mu\nu}]
  &=
  2i \Big(
    \sigma^{\pi\mu} g^{\rho\nu}
    + \sigma^{\rho\nu} g^{\pi\mu}
    - \sigma^{\pi\nu} g^{\rho\mu}
    - \sigma^{\rho\mu} g^{\pi\nu}
    \Big)
  \,,
\end{align}
we find:
\begin{align}
  [\hat{M},\sigma^{\mu\nu}]
  &=
  -G_\omega
  \overline\psi \gamma^{[\mu} \psi \gamma^{\nu]}
  -G_f
  \overline\psi \gamma^{[\mu}\gamma_5 \psi \gamma^{\nu]} \gamma_5
  -G_\rho \left(
    \overline\psi \gamma^{[\mu} \tau_i \psi \gamma^{\nu]} \tau_i
    +
    \overline\psi \gamma^{[\mu}\gamma_5 \tau_i \psi \gamma^{\nu]}\gamma_5 \tau_i
    \right)
  \notag \\ &
  - 2 G_T \Big(
    (\overline\psi i\sigma^{\pi[\mu}\psi)
    - (\overline{\psi}i\sigma^{\pi[\mu}\tau_i\psi) \tau_i
    \Big) \sigma^{\nu]\pi}
\end{align}
Finally, sandwiching this between $\psi$ and $\overline{\psi}$,
we get
\begin{align}
  \overline\psi [\hat{M},\sigma^{\mu\nu}] \psi
  &=
  -G_\omega
  \overline\psi \gamma^{[\mu} \psi
  \overline{\psi} \gamma^{\nu]} \psi
  -
  G_f
  \overline\psi \gamma^{[\mu}\gamma_5 \psi
  \overline{\psi} \gamma^{\nu]} \gamma_5 \psi
  -
  G_\rho \left(
    \overline\psi \gamma^{[\mu} \tau_i \psi
    \overline\psi \gamma^{\nu]} \tau_i \psi
    +
    \overline\psi \gamma^{[\mu}\gamma_5 \tau_i \psi
    \overline{\psi} \gamma^{\nu]}\gamma_5 \tau_i \psi
    \right)
  \notag \\ &
  - 2 G_T \Big(
    (\overline\psi i\sigma^{\pi[\mu}\psi)
    (\overline\psi i\sigma^{\nu]\pi}\psi)
    -
    (\overline\psi i\sigma^{\pi[\mu}\tau_i\psi)
    (\overline\psi i\sigma^{\nu]\pi}\tau_i\psi)
    \Big)
  \,,
\end{align}
an expression which is clearly both symmetric and antisymmetric under the
swap $(\mu\leftrightarrow\nu)$, and which is therefore zero.
We have thus proved Eq.~(\ref{eqn:eom:5}) holds in the NJL model.

\bibliography{main.bib}

\begin{thebibliography}{51}%
\makeatletter
\providecommand \@ifxundefined [1]{%
 \@ifx{#1\undefined}
}%
\providecommand \@ifnum [1]{%
 \ifnum #1\expandafter \@firstoftwo
 \else \expandafter \@secondoftwo
 \fi
}%
\providecommand \@ifx [1]{%
 \ifx #1\expandafter \@firstoftwo
 \else \expandafter \@secondoftwo
 \fi
}%
\providecommand \natexlab [1]{#1}%
\providecommand \enquote  [1]{``#1''}%
\providecommand \bibnamefont  [1]{#1}%
\providecommand \bibfnamefont [1]{#1}%
\providecommand \citenamefont [1]{#1}%
\providecommand \href@noop [0]{\@secondoftwo}%
\providecommand \href [0]{\begingroup \@sanitize@url \@href}%
\providecommand \@href[1]{\@@startlink{#1}\@@href}%
\providecommand \@@href[1]{\endgroup#1\@@endlink}%
\providecommand \@sanitize@url [0]{\catcode `\\12\catcode `\$12\catcode
  `\&12\catcode `\#12\catcode `\^12\catcode `\_12\catcode `\%12\relax}%
\providecommand \@@startlink[1]{}%
\providecommand \@@endlink[0]{}%
\providecommand \url  [0]{\begingroup\@sanitize@url \@url }%
\providecommand \@url [1]{\endgroup\@href {#1}{\urlprefix }}%
\providecommand \urlprefix  [0]{URL }%
\providecommand \Eprint [0]{\href }%
\providecommand \doibase [0]{http://dx.doi.org/}%
\providecommand \selectlanguage [0]{\@gobble}%
\providecommand \bibinfo  [0]{\@secondoftwo}%
\providecommand \bibfield  [0]{\@secondoftwo}%
\providecommand \translation [1]{[#1]}%
\providecommand \BibitemOpen [0]{}%
\providecommand \bibitemStop [0]{}%
\providecommand \bibitemNoStop [0]{.\EOS\space}%
\providecommand \EOS [0]{\spacefactor3000\relax}%
\providecommand \BibitemShut  [1]{\csname bibitem#1\endcsname}%
\let\auto@bib@innerbib\@empty
\bibitem [{\citenamefont {Leader}\ and\ \citenamefont
  {Lorcé}(2014)}]{Leader:2013jra}%
  \BibitemOpen
  \bibfield  {author} {\bibinfo {author} {\bibfnamefont {E.}~\bibnamefont
  {Leader}}\ and\ \bibinfo {author} {\bibfnamefont {C.}~\bibnamefont
  {Lorcé}},\ }\href {\doibase 10.1016/j.physrep.2014.02.010} {\bibfield
  {journal} {\bibinfo  {journal} {Phys. Rept.}\ }\textbf {\bibinfo {volume}
  {541}},\ \bibinfo {pages} {163} (\bibinfo {year} {2014})},\ \Eprint
  {http://arxiv.org/abs/1309.4235} {arXiv:1309.4235 [hep-ph]} \BibitemShut
  {NoStop}%
\bibitem [{\citenamefont {Polyakov}\ and\ \citenamefont
  {Schweitzer}(2018)}]{Polyakov:2018zvc}%
  \BibitemOpen
  \bibfield  {author} {\bibinfo {author} {\bibfnamefont {M.~V.}\ \bibnamefont
  {Polyakov}}\ and\ \bibinfo {author} {\bibfnamefont {P.}~\bibnamefont
  {Schweitzer}},\ }\href {\doibase 10.1142/S0217751X18300259} {\bibfield
  {journal} {\bibinfo  {journal} {Int. J. Mod. Phys.}\ }\textbf {\bibinfo
  {volume} {A33}},\ \bibinfo {pages} {1830025} (\bibinfo {year} {2018})},\
  \Eprint {http://arxiv.org/abs/1805.06596} {arXiv:1805.06596 [hep-ph]}
  \BibitemShut {NoStop}%
\bibitem [{\citenamefont {Polyakov}(2003)}]{Polyakov:2002yz}%
  \BibitemOpen
  \bibfield  {author} {\bibinfo {author} {\bibfnamefont {M.~V.}\ \bibnamefont
  {Polyakov}},\ }\href {\doibase 10.1016/S0370-2693(03)00036-4} {\bibfield
  {journal} {\bibinfo  {journal} {Phys. Lett.}\ }\textbf {\bibinfo {volume}
  {B555}},\ \bibinfo {pages} {57} (\bibinfo {year} {2003})},\ \Eprint
  {http://arxiv.org/abs/hep-ph/0210165} {arXiv:hep-ph/0210165 [hep-ph]}
  \BibitemShut {NoStop}%
\bibitem [{\citenamefont {Lorcé}(2018{\natexlab{a}})}]{Lorce:2017xzd}%
  \BibitemOpen
  \bibfield  {author} {\bibinfo {author} {\bibfnamefont {C.}~\bibnamefont
  {Lorcé}},\ }\href {\doibase 10.1140/epjc/s10052-018-5561-2} {\bibfield
  {journal} {\bibinfo  {journal} {Eur. Phys. J.}\ }\textbf {\bibinfo {volume}
  {C78}},\ \bibinfo {pages} {120} (\bibinfo {year} {2018}{\natexlab{a}})},\
  \Eprint {http://arxiv.org/abs/1706.05853} {arXiv:1706.05853 [hep-ph]}
  \BibitemShut {NoStop}%
\bibitem [{\citenamefont {Lorcé}\ \emph {et~al.}(2018)\citenamefont {Lorcé},
  \citenamefont {Mantovani},\ and\ \citenamefont {Pasquini}}]{Lorce:2017wkb}%
  \BibitemOpen
  \bibfield  {author} {\bibinfo {author} {\bibfnamefont {C.}~\bibnamefont
  {Lorcé}}, \bibinfo {author} {\bibfnamefont {L.}~\bibnamefont {Mantovani}}, \
  and\ \bibinfo {author} {\bibfnamefont {B.}~\bibnamefont {Pasquini}},\ }\href
  {\doibase 10.1016/j.physletb.2017.11.018} {\bibfield  {journal} {\bibinfo
  {journal} {Phys. Lett.}\ }\textbf {\bibinfo {volume} {B776}},\ \bibinfo
  {pages} {38} (\bibinfo {year} {2018})},\ \Eprint
  {http://arxiv.org/abs/1704.08557} {arXiv:1704.08557 [hep-ph]} \BibitemShut
  {NoStop}%
\bibitem [{\citenamefont {Polyakov}\ and\ \citenamefont
  {Shuvaev}(2002)}]{Polyakov:2002wz}%
  \BibitemOpen
  \bibfield  {author} {\bibinfo {author} {\bibfnamefont {M.~V.}\ \bibnamefont
  {Polyakov}}\ and\ \bibinfo {author} {\bibfnamefont {A.~G.}\ \bibnamefont
  {Shuvaev}},\ }\href@noop {} {\  (\bibinfo {year} {2002})},\ \Eprint
  {http://arxiv.org/abs/hep-ph/0207153} {arXiv:hep-ph/0207153 [hep-ph]}
  \BibitemShut {NoStop}%
\bibitem [{\citenamefont {Lorcé}\ \emph {et~al.}(2019)\citenamefont {Lorcé},
  \citenamefont {Moutarde},\ and\ \citenamefont {Trawiński}}]{Lorce:2018egm}%
  \BibitemOpen
  \bibfield  {author} {\bibinfo {author} {\bibfnamefont {C.}~\bibnamefont
  {Lorcé}}, \bibinfo {author} {\bibfnamefont {H.}~\bibnamefont {Moutarde}}, \
  and\ \bibinfo {author} {\bibfnamefont {A.~P.}\ \bibnamefont {Trawiński}},\
  }\href {\doibase 10.1140/epjc/s10052-019-6572-3} {\bibfield  {journal}
  {\bibinfo  {journal} {Eur. Phys. J.}\ }\textbf {\bibinfo {volume} {C79}},\
  \bibinfo {pages} {89} (\bibinfo {year} {2019})},\ \Eprint
  {http://arxiv.org/abs/1810.09837} {arXiv:1810.09837 [hep-ph]} \BibitemShut
  {NoStop}%
\bibitem [{\citenamefont {Novikov}\ and\ \citenamefont
  {Shifman}(1981)}]{Novikov:1980fa}%
  \BibitemOpen
  \bibfield  {author} {\bibinfo {author} {\bibfnamefont {V.~A.}\ \bibnamefont
  {Novikov}}\ and\ \bibinfo {author} {\bibfnamefont {M.~A.}\ \bibnamefont
  {Shifman}},\ }\href {\doibase 10.1007/BF01429829} {\bibfield  {journal}
  {\bibinfo  {journal} {Z. Phys.}\ }\textbf {\bibinfo {volume} {C8}},\ \bibinfo
  {pages} {43} (\bibinfo {year} {1981})}\BibitemShut {NoStop}%
\bibitem [{\citenamefont {Voloshin}\ and\ \citenamefont
  {Zakharov}(1980)}]{Voloshin:1980zf}%
  \BibitemOpen
  \bibfield  {author} {\bibinfo {author} {\bibfnamefont {M.~B.}\ \bibnamefont
  {Voloshin}}\ and\ \bibinfo {author} {\bibfnamefont {V.~I.}\ \bibnamefont
  {Zakharov}},\ }\href {\doibase 10.1103/PhysRevLett.45.688} {\bibfield
  {journal} {\bibinfo  {journal} {Phys. Rev. Lett.}\ }\textbf {\bibinfo
  {volume} {45}},\ \bibinfo {pages} {688} (\bibinfo {year} {1980})}\BibitemShut
  {NoStop}%
\bibitem [{\citenamefont {Polyakov}\ and\ \citenamefont
  {Weiss}(1999)}]{Polyakov:1999gs}%
  \BibitemOpen
  \bibfield  {author} {\bibinfo {author} {\bibfnamefont {M.~V.}\ \bibnamefont
  {Polyakov}}\ and\ \bibinfo {author} {\bibfnamefont {C.}~\bibnamefont
  {Weiss}},\ }\href {\doibase 10.1103/PhysRevD.60.114017} {\bibfield  {journal}
  {\bibinfo  {journal} {Phys. Rev.}\ }\textbf {\bibinfo {volume} {D60}},\
  \bibinfo {pages} {114017} (\bibinfo {year} {1999})},\ \Eprint
  {http://arxiv.org/abs/hep-ph/9902451} {arXiv:hep-ph/9902451 [hep-ph]}
  \BibitemShut {NoStop}%
\bibitem [{\citenamefont {Nambu}\ and\ \citenamefont
  {Jona-Lasinio}(1961{\natexlab{a}})}]{Nambu:1961tp}%
  \BibitemOpen
  \bibfield  {author} {\bibinfo {author} {\bibfnamefont {Y.}~\bibnamefont
  {Nambu}}\ and\ \bibinfo {author} {\bibfnamefont {G.}~\bibnamefont
  {Jona-Lasinio}},\ }\href {\doibase 10.1103/PhysRev.122.345} {\bibfield
  {journal} {\bibinfo  {journal} {Phys. Rev.}\ }\textbf {\bibinfo {volume}
  {122}},\ \bibinfo {pages} {345} (\bibinfo {year} {1961}{\natexlab{a}})},\
  \bibinfo {note} {[,127(1961)]}\BibitemShut {NoStop}%
\bibitem [{\citenamefont {Nambu}\ and\ \citenamefont
  {Jona-Lasinio}(1961{\natexlab{b}})}]{Nambu:1961fr}%
  \BibitemOpen
  \bibfield  {author} {\bibinfo {author} {\bibfnamefont {Y.}~\bibnamefont
  {Nambu}}\ and\ \bibinfo {author} {\bibfnamefont {G.}~\bibnamefont
  {Jona-Lasinio}},\ }\href {\doibase 10.1103/PhysRev.124.246} {\bibfield
  {journal} {\bibinfo  {journal} {Phys. Rev.}\ }\textbf {\bibinfo {volume}
  {124}},\ \bibinfo {pages} {246} (\bibinfo {year} {1961}{\natexlab{b}})},\
  \bibinfo {note} {[,141(1961)]}\BibitemShut {NoStop}%
\bibitem [{\citenamefont {Bashir}\ \emph {et~al.}(2012)\citenamefont {Bashir},
  \citenamefont {Chang}, \citenamefont {Cloet}, \citenamefont {El-Bennich},
  \citenamefont {Liu}, \citenamefont {Roberts},\ and\ \citenamefont
  {Tandy}}]{Bashir:2012fs}%
  \BibitemOpen
  \bibfield  {author} {\bibinfo {author} {\bibfnamefont {A.}~\bibnamefont
  {Bashir}}, \bibinfo {author} {\bibfnamefont {L.}~\bibnamefont {Chang}},
  \bibinfo {author} {\bibfnamefont {I.~C.}\ \bibnamefont {Cloet}}, \bibinfo
  {author} {\bibfnamefont {B.}~\bibnamefont {El-Bennich}}, \bibinfo {author}
  {\bibfnamefont {Y.-X.}\ \bibnamefont {Liu}}, \bibinfo {author} {\bibfnamefont
  {C.~D.}\ \bibnamefont {Roberts}}, \ and\ \bibinfo {author} {\bibfnamefont
  {P.~C.}\ \bibnamefont {Tandy}},\ }\href {\doibase 10.1088/0253-6102/58/1/16}
  {\bibfield  {journal} {\bibinfo  {journal} {Commun. Theor. Phys.}\ }\textbf
  {\bibinfo {volume} {58}},\ \bibinfo {pages} {79} (\bibinfo {year} {2012})},\
  \Eprint {http://arxiv.org/abs/1201.3366} {arXiv:1201.3366 [nucl-th]}
  \BibitemShut {NoStop}%
\bibitem [{\citenamefont {Brommel}(2007)}]{Brommel:2007zz}%
  \BibitemOpen
  \bibfield  {author} {\bibinfo {author} {\bibfnamefont {D.}~\bibnamefont
  {Brommel}},\ }\emph {\bibinfo {title} {{Pion Structure from the Lattice}}},\
  \href {\doibase 10.3204/DESY-THESIS-2007-023} {Ph.D. thesis},\ \bibinfo
  {school} {Regensburg U.} (\bibinfo {year} {2007})\BibitemShut {NoStop}%
\bibitem [{\citenamefont {Shanahan}\ and\ \citenamefont
  {Detmold}(2019)}]{Shanahan:2018pib}%
  \BibitemOpen
  \bibfield  {author} {\bibinfo {author} {\bibfnamefont {P.~E.}\ \bibnamefont
  {Shanahan}}\ and\ \bibinfo {author} {\bibfnamefont {W.}~\bibnamefont
  {Detmold}},\ }\href {\doibase 10.1103/PhysRevD.99.014511} {\bibfield
  {journal} {\bibinfo  {journal} {Phys. Rev.}\ }\textbf {\bibinfo {volume}
  {D99}},\ \bibinfo {pages} {014511} (\bibinfo {year} {2019})},\ \Eprint
  {http://arxiv.org/abs/1810.04626} {arXiv:1810.04626 [hep-lat]} \BibitemShut
  {NoStop}%
\bibitem [{\citenamefont {Kumano}\ \emph {et~al.}(2018)\citenamefont {Kumano},
  \citenamefont {Song},\ and\ \citenamefont {Teryaev}}]{Kumano:2017lhr}%
  \BibitemOpen
  \bibfield  {author} {\bibinfo {author} {\bibfnamefont {S.}~\bibnamefont
  {Kumano}}, \bibinfo {author} {\bibfnamefont {Q.-T.}\ \bibnamefont {Song}}, \
  and\ \bibinfo {author} {\bibfnamefont {O.~V.}\ \bibnamefont {Teryaev}},\
  }\href {\doibase 10.1103/PhysRevD.97.014020} {\bibfield  {journal} {\bibinfo
  {journal} {Phys. Rev.}\ }\textbf {\bibinfo {volume} {D97}},\ \bibinfo {pages}
  {014020} (\bibinfo {year} {2018})},\ \Eprint
  {http://arxiv.org/abs/1711.08088} {arXiv:1711.08088 [hep-ph]} \BibitemShut
  {NoStop}%
\bibitem [{\citenamefont {Vogl}\ and\ \citenamefont
  {Weise}(1991)}]{Vogl:1991qt}%
  \BibitemOpen
  \bibfield  {author} {\bibinfo {author} {\bibfnamefont {U.}~\bibnamefont
  {Vogl}}\ and\ \bibinfo {author} {\bibfnamefont {W.}~\bibnamefont {Weise}},\
  }\href {\doibase 10.1016/0146-6410(91)90005-9} {\bibfield  {journal}
  {\bibinfo  {journal} {Prog. Part. Nucl. Phys.}\ }\textbf {\bibinfo {volume}
  {27}},\ \bibinfo {pages} {195} (\bibinfo {year} {1991})}\BibitemShut
  {NoStop}%
\bibitem [{\citenamefont {Klevansky}(1992)}]{Klevansky:1992qe}%
  \BibitemOpen
  \bibfield  {author} {\bibinfo {author} {\bibfnamefont {S.~P.}\ \bibnamefont
  {Klevansky}},\ }\href {\doibase 10.1103/RevModPhys.64.649} {\bibfield
  {journal} {\bibinfo  {journal} {Rev. Mod. Phys.}\ }\textbf {\bibinfo {volume}
  {64}},\ \bibinfo {pages} {649} (\bibinfo {year} {1992})}\BibitemShut
  {NoStop}%
\bibitem [{\citenamefont {Hatsuda}\ and\ \citenamefont
  {Kunihiro}(1994)}]{Hatsuda:1994pi}%
  \BibitemOpen
  \bibfield  {author} {\bibinfo {author} {\bibfnamefont {T.}~\bibnamefont
  {Hatsuda}}\ and\ \bibinfo {author} {\bibfnamefont {T.}~\bibnamefont
  {Kunihiro}},\ }\href {\doibase 10.1016/0370-1573(94)90022-1} {\bibfield
  {journal} {\bibinfo  {journal} {Phys. Rept.}\ }\textbf {\bibinfo {volume}
  {247}},\ \bibinfo {pages} {221} (\bibinfo {year} {1994})},\ \Eprint
  {http://arxiv.org/abs/hep-ph/9401310} {arXiv:hep-ph/9401310 [hep-ph]}
  \BibitemShut {NoStop}%
\bibitem [{\citenamefont {Cloët}\ \emph {et~al.}(2014)\citenamefont {Cloët},
  \citenamefont {Bentz},\ and\ \citenamefont {Thomas}}]{Cloet:2014rja}%
  \BibitemOpen
  \bibfield  {author} {\bibinfo {author} {\bibfnamefont {I.~C.}\ \bibnamefont
  {Cloët}}, \bibinfo {author} {\bibfnamefont {W.}~\bibnamefont {Bentz}}, \
  and\ \bibinfo {author} {\bibfnamefont {A.~W.}\ \bibnamefont {Thomas}},\
  }\href {\doibase 10.1103/PhysRevC.90.045202} {\bibfield  {journal} {\bibinfo
  {journal} {Phys. Rev.}\ }\textbf {\bibinfo {volume} {C90}},\ \bibinfo {pages}
  {045202} (\bibinfo {year} {2014})},\ \Eprint {http://arxiv.org/abs/1405.5542}
  {arXiv:1405.5542 [nucl-th]} \BibitemShut {NoStop}%
\bibitem [{\citenamefont {Ninomiya}\ \emph {et~al.}(2017)\citenamefont
  {Ninomiya}, \citenamefont {Bentz},\ and\ \citenamefont
  {Cloët}}]{Ninomiya:2017ggn}%
  \BibitemOpen
  \bibfield  {author} {\bibinfo {author} {\bibfnamefont {Y.}~\bibnamefont
  {Ninomiya}}, \bibinfo {author} {\bibfnamefont {W.}~\bibnamefont {Bentz}}, \
  and\ \bibinfo {author} {\bibfnamefont {I.~C.}\ \bibnamefont {Cloët}},\
  }\href {\doibase 10.1103/PhysRevC.96.045206} {\bibfield  {journal} {\bibinfo
  {journal} {Phys. Rev.}\ }\textbf {\bibinfo {volume} {C96}},\ \bibinfo {pages}
  {045206} (\bibinfo {year} {2017})},\ \Eprint
  {http://arxiv.org/abs/1707.03787} {arXiv:1707.03787 [nucl-th]} \BibitemShut
  {NoStop}%
\bibitem [{\citenamefont {Ishii}\ \emph
  {et~al.}(1993{\natexlab{a}})\citenamefont {Ishii}, \citenamefont {Bentz},\
  and\ \citenamefont {Yazaki}}]{Ishii:1993np}%
  \BibitemOpen
  \bibfield  {author} {\bibinfo {author} {\bibfnamefont {N.}~\bibnamefont
  {Ishii}}, \bibinfo {author} {\bibfnamefont {W.}~\bibnamefont {Bentz}}, \ and\
  \bibinfo {author} {\bibfnamefont {K.}~\bibnamefont {Yazaki}},\ }\href
  {\doibase 10.1016/0370-2693(93)90683-9} {\bibfield  {journal} {\bibinfo
  {journal} {Phys. Lett.}\ }\textbf {\bibinfo {volume} {B301}},\ \bibinfo
  {pages} {165} (\bibinfo {year} {1993}{\natexlab{a}})}\BibitemShut {NoStop}%
\bibitem [{\citenamefont {Ishii}\ \emph
  {et~al.}(1993{\natexlab{b}})\citenamefont {Ishii}, \citenamefont {Bentz},\
  and\ \citenamefont {Yazaki}}]{Ishii:1993rt}%
  \BibitemOpen
  \bibfield  {author} {\bibinfo {author} {\bibfnamefont {N.}~\bibnamefont
  {Ishii}}, \bibinfo {author} {\bibfnamefont {W.}~\bibnamefont {Bentz}}, \ and\
  \bibinfo {author} {\bibfnamefont {K.}~\bibnamefont {Yazaki}},\ }\href
  {\doibase 10.1016/0370-2693(93)91778-L} {\bibfield  {journal} {\bibinfo
  {journal} {Phys. Lett.}\ }\textbf {\bibinfo {volume} {B318}},\ \bibinfo
  {pages} {26} (\bibinfo {year} {1993}{\natexlab{b}})}\BibitemShut {NoStop}%
\bibitem [{\citenamefont {Ishii}\ \emph {et~al.}(1995)\citenamefont {Ishii},
  \citenamefont {Bentz},\ and\ \citenamefont {Yazaki}}]{Ishii:1995bu}%
  \BibitemOpen
  \bibfield  {author} {\bibinfo {author} {\bibfnamefont {N.}~\bibnamefont
  {Ishii}}, \bibinfo {author} {\bibfnamefont {W.}~\bibnamefont {Bentz}}, \ and\
  \bibinfo {author} {\bibfnamefont {K.}~\bibnamefont {Yazaki}},\ }\href
  {\doibase 10.1016/0375-9474(95)00032-V} {\bibfield  {journal} {\bibinfo
  {journal} {Nucl. Phys.}\ }\textbf {\bibinfo {volume} {A587}},\ \bibinfo
  {pages} {617} (\bibinfo {year} {1995})}\BibitemShut {NoStop}%
\bibitem [{\citenamefont {Ebert}\ \emph {et~al.}(1996)\citenamefont {Ebert},
  \citenamefont {Feldmann},\ and\ \citenamefont {Reinhardt}}]{Ebert:1996vx}%
  \BibitemOpen
  \bibfield  {author} {\bibinfo {author} {\bibfnamefont {D.}~\bibnamefont
  {Ebert}}, \bibinfo {author} {\bibfnamefont {T.}~\bibnamefont {Feldmann}}, \
  and\ \bibinfo {author} {\bibfnamefont {H.}~\bibnamefont {Reinhardt}},\ }\href
  {\doibase 10.1016/0370-2693(96)01158-6} {\bibfield  {journal} {\bibinfo
  {journal} {Phys. Lett.}\ }\textbf {\bibinfo {volume} {B388}},\ \bibinfo
  {pages} {154} (\bibinfo {year} {1996})},\ \Eprint
  {http://arxiv.org/abs/hep-ph/9608223} {arXiv:hep-ph/9608223 [hep-ph]}
  \BibitemShut {NoStop}%
\bibitem [{\citenamefont {Hellstern}\ \emph {et~al.}(1997)\citenamefont
  {Hellstern}, \citenamefont {Alkofer},\ and\ \citenamefont
  {Reinhardt}}]{Hellstern:1997nv}%
  \BibitemOpen
  \bibfield  {author} {\bibinfo {author} {\bibfnamefont {G.}~\bibnamefont
  {Hellstern}}, \bibinfo {author} {\bibfnamefont {R.}~\bibnamefont {Alkofer}},
  \ and\ \bibinfo {author} {\bibfnamefont {H.}~\bibnamefont {Reinhardt}},\
  }\href {\doibase 10.1016/S0375-9474(97)00412-0} {\bibfield  {journal}
  {\bibinfo  {journal} {Nucl. Phys.}\ }\textbf {\bibinfo {volume} {A625}},\
  \bibinfo {pages} {697} (\bibinfo {year} {1997})},\ \Eprint
  {http://arxiv.org/abs/hep-ph/9706551} {arXiv:hep-ph/9706551 [hep-ph]}
  \BibitemShut {NoStop}%
\bibitem [{\citenamefont {Cartan}(1923)}]{Cartan:1923zea}%
  \BibitemOpen
  \bibfield  {author} {\bibinfo {author} {\bibfnamefont {E.}~\bibnamefont
  {Cartan}},\ }\href@noop {} {\bibfield  {journal} {\bibinfo  {journal}
  {Annales Sci. Ecole Norm. Sup.}\ }\textbf {\bibinfo {volume} {40}},\ \bibinfo
  {pages} {325} (\bibinfo {year} {1923})}\BibitemShut {NoStop}%
\bibitem [{\citenamefont {Cartan}(1924)}]{Cartan:1924yea}%
  \BibitemOpen
  \bibfield  {author} {\bibinfo {author} {\bibfnamefont {E.}~\bibnamefont
  {Cartan}},\ }\href@noop {} {\bibfield  {journal} {\bibinfo  {journal}
  {Annales Sci. Ecole Norm. Sup.}\ }\textbf {\bibinfo {volume} {41}},\ \bibinfo
  {pages} {1} (\bibinfo {year} {1924})}\BibitemShut {NoStop}%
\bibitem [{\citenamefont {Brout}\ and\ \citenamefont
  {Englert}(1966)}]{Brout:1966oea}%
  \BibitemOpen
  \bibfield  {author} {\bibinfo {author} {\bibfnamefont {R.}~\bibnamefont
  {Brout}}\ and\ \bibinfo {author} {\bibfnamefont {F.}~\bibnamefont
  {Englert}},\ }\href {\doibase 10.1103/PhysRev.141.1231} {\bibfield  {journal}
  {\bibinfo  {journal} {Phys. Rev.}\ }\textbf {\bibinfo {volume} {141}},\
  \bibinfo {pages} {1231} (\bibinfo {year} {1966})}\BibitemShut {NoStop}%
\bibitem [{\citenamefont {Lorcé}(2018{\natexlab{b}})}]{Lorce:2018zpf}%
  \BibitemOpen
  \bibfield  {author} {\bibinfo {author} {\bibfnamefont {C.}~\bibnamefont
  {Lorcé}},\ }\href {\doibase 10.1140/epjc/s10052-018-6249-3} {\bibfield
  {journal} {\bibinfo  {journal} {Eur. Phys. J.}\ }\textbf {\bibinfo {volume}
  {C78}},\ \bibinfo {pages} {785} (\bibinfo {year} {2018}{\natexlab{b}})},\
  \Eprint {http://arxiv.org/abs/1805.05284} {arXiv:1805.05284 [hep-ph]}
  \BibitemShut {NoStop}%
\bibitem [{\citenamefont {Florkowski}\ and\ \citenamefont
  {Ryblewski}(2018)}]{Florkowski:2018fap}%
  \BibitemOpen
  \bibfield  {author} {\bibinfo {author} {\bibfnamefont {W.}~\bibnamefont
  {Florkowski}}\ and\ \bibinfo {author} {\bibfnamefont {R.}~\bibnamefont
  {Ryblewski}},\ }\href@noop {} {\  (\bibinfo {year} {2018})},\ \Eprint
  {http://arxiv.org/abs/1811.04409} {arXiv:1811.04409 [nucl-th]} \BibitemShut
  {NoStop}%
\bibitem [{\citenamefont {Hudson}\ and\ \citenamefont
  {Schweitzer}(2017)}]{Hudson:2017xug}%
  \BibitemOpen
  \bibfield  {author} {\bibinfo {author} {\bibfnamefont {J.}~\bibnamefont
  {Hudson}}\ and\ \bibinfo {author} {\bibfnamefont {P.}~\bibnamefont
  {Schweitzer}},\ }\href {\doibase 10.1103/PhysRevD.96.114013} {\bibfield
  {journal} {\bibinfo  {journal} {Phys. Rev.}\ }\textbf {\bibinfo {volume}
  {D96}},\ \bibinfo {pages} {114013} (\bibinfo {year} {2017})},\ \Eprint
  {http://arxiv.org/abs/1712.05316} {arXiv:1712.05316 [hep-ph]} \BibitemShut
  {NoStop}%
\bibitem [{\citenamefont {Ji}(1997)}]{Ji:1996nm}%
  \BibitemOpen
  \bibfield  {author} {\bibinfo {author} {\bibfnamefont {X.-D.}\ \bibnamefont
  {Ji}},\ }\href {\doibase 10.1103/PhysRevD.55.7114} {\bibfield  {journal}
  {\bibinfo  {journal} {Phys. Rev.}\ }\textbf {\bibinfo {volume} {D55}},\
  \bibinfo {pages} {7114} (\bibinfo {year} {1997})},\ \Eprint
  {http://arxiv.org/abs/hep-ph/9609381} {arXiv:hep-ph/9609381 [hep-ph]}
  \BibitemShut {NoStop}%
\bibitem [{\citenamefont {DeWitt}(1967)}]{DeWitt:1967uc}%
  \BibitemOpen
  \bibfield  {author} {\bibinfo {author} {\bibfnamefont {B.~S.}\ \bibnamefont
  {DeWitt}},\ }\href {\doibase 10.1103/PhysRev.162.1239} {\bibfield  {journal}
  {\bibinfo  {journal} {Phys. Rev.}\ }\textbf {\bibinfo {volume} {162}},\
  \bibinfo {pages} {1239} (\bibinfo {year} {1967})},\ \bibinfo {note}
  {[,307(1967)]}\BibitemShut {NoStop}%
\bibitem [{\citenamefont {Bessler}\ \emph {et~al.}(1969)\citenamefont
  {Bessler}, \citenamefont {Muta},\ and\ \citenamefont
  {Umezawa}}]{Bessler:1969py}%
  \BibitemOpen
  \bibfield  {author} {\bibinfo {author} {\bibfnamefont {L.}~\bibnamefont
  {Bessler}}, \bibinfo {author} {\bibfnamefont {T.}~\bibnamefont {Muta}}, \
  and\ \bibinfo {author} {\bibfnamefont {H.}~\bibnamefont {Umezawa}},\ }\href
  {\doibase 10.1103/PhysRev.180.1604} {\bibfield  {journal} {\bibinfo
  {journal} {Phys. Rev.}\ }\textbf {\bibinfo {volume} {180}},\ \bibinfo {pages}
  {1604} (\bibinfo {year} {1969})}\BibitemShut {NoStop}%
\bibitem [{\citenamefont {Cosyn}\ \emph {et~al.}(2019)\citenamefont {Cosyn},
  \citenamefont {Cotogno}, \citenamefont {Freese},\ and\ \citenamefont
  {Lorcé}}]{Cosyn:2019aio}%
  \BibitemOpen
  \bibfield  {author} {\bibinfo {author} {\bibfnamefont {W.}~\bibnamefont
  {Cosyn}}, \bibinfo {author} {\bibfnamefont {S.}~\bibnamefont {Cotogno}},
  \bibinfo {author} {\bibfnamefont {A.}~\bibnamefont {Freese}}, \ and\ \bibinfo
  {author} {\bibfnamefont {C.}~\bibnamefont {Lorcé}},\ }\href@noop {} {\
  (\bibinfo {year} {2019})},\ \Eprint {http://arxiv.org/abs/1903.00408}
  {arXiv:1903.00408 [hep-ph]} \BibitemShut {NoStop}%
\bibitem [{\citenamefont {Polyakov}\ and\ \citenamefont
  {Sun}(2019)}]{Polyakov:2019lbq}%
  \BibitemOpen
  \bibfield  {author} {\bibinfo {author} {\bibfnamefont {M.~V.}\ \bibnamefont
  {Polyakov}}\ and\ \bibinfo {author} {\bibfnamefont {B.-D.}\ \bibnamefont
  {Sun}},\ }\href@noop {} {\  (\bibinfo {year} {2019})},\ \Eprint
  {http://arxiv.org/abs/1903.02738} {arXiv:1903.02738 [hep-ph]} \BibitemShut
  {NoStop}%
\bibitem [{\citenamefont {Hudson}\ and\ \citenamefont
  {Schweitzer}(2018)}]{Hudson:2017oul}%
  \BibitemOpen
  \bibfield  {author} {\bibinfo {author} {\bibfnamefont {J.}~\bibnamefont
  {Hudson}}\ and\ \bibinfo {author} {\bibfnamefont {P.}~\bibnamefont
  {Schweitzer}},\ }\href {\doibase 10.1103/PhysRevD.97.056003} {\bibfield
  {journal} {\bibinfo  {journal} {Phys. Rev.}\ }\textbf {\bibinfo {volume}
  {D97}},\ \bibinfo {pages} {056003} (\bibinfo {year} {2018})},\ \Eprint
  {http://arxiv.org/abs/1712.05317} {arXiv:1712.05317 [hep-ph]} \BibitemShut
  {NoStop}%
\bibitem [{\citenamefont {Frederico}\ \emph {et~al.}(1992)\citenamefont
  {Frederico}, \citenamefont {Henley}, \citenamefont {Pollock},\ and\
  \citenamefont {Ying}}]{Frederico:1992vm}%
  \BibitemOpen
  \bibfield  {author} {\bibinfo {author} {\bibfnamefont {T.}~\bibnamefont
  {Frederico}}, \bibinfo {author} {\bibfnamefont {E.~M.}\ \bibnamefont
  {Henley}}, \bibinfo {author} {\bibfnamefont {S.~J.}\ \bibnamefont {Pollock}},
  \ and\ \bibinfo {author} {\bibfnamefont {S.}~\bibnamefont {Ying}},\ }\href
  {\doibase 10.1103/PhysRevC.46.347} {\bibfield  {journal} {\bibinfo  {journal}
  {Phys. Rev.}\ }\textbf {\bibinfo {volume} {C46}},\ \bibinfo {pages} {347}
  (\bibinfo {year} {1992})}\BibitemShut {NoStop}%
\bibitem [{\citenamefont {Berger}\ \emph {et~al.}(2001)\citenamefont {Berger},
  \citenamefont {Cano}, \citenamefont {Diehl},\ and\ \citenamefont
  {Pire}}]{Berger:2001zb}%
  \BibitemOpen
  \bibfield  {author} {\bibinfo {author} {\bibfnamefont {E.~R.}\ \bibnamefont
  {Berger}}, \bibinfo {author} {\bibfnamefont {F.}~\bibnamefont {Cano}},
  \bibinfo {author} {\bibfnamefont {M.}~\bibnamefont {Diehl}}, \ and\ \bibinfo
  {author} {\bibfnamefont {B.}~\bibnamefont {Pire}},\ }\href {\doibase
  10.1103/PhysRevLett.87.142302} {\bibfield  {journal} {\bibinfo  {journal}
  {Phys. Rev. Lett.}\ }\textbf {\bibinfo {volume} {87}},\ \bibinfo {pages}
  {142302} (\bibinfo {year} {2001})},\ \Eprint
  {http://arxiv.org/abs/hep-ph/0106192} {arXiv:hep-ph/0106192 [hep-ph]}
  \BibitemShut {NoStop}%
\bibitem [{\citenamefont {Adler}(1969)}]{Adler:1969gk}%
  \BibitemOpen
  \bibfield  {author} {\bibinfo {author} {\bibfnamefont {S.~L.}\ \bibnamefont
  {Adler}},\ }\href {\doibase 10.1103/PhysRev.177.2426} {\bibfield  {journal}
  {\bibinfo  {journal} {Phys. Rev.}\ }\textbf {\bibinfo {volume} {177}},\
  \bibinfo {pages} {2426} (\bibinfo {year} {1969})},\ \bibinfo {note}
  {[,241(1969)]}\BibitemShut {NoStop}%
\bibitem [{\citenamefont {Ji}(1998)}]{Ji:1998pc}%
  \BibitemOpen
  \bibfield  {author} {\bibinfo {author} {\bibfnamefont {X.-D.}\ \bibnamefont
  {Ji}},\ }\href {\doibase 10.1088/0954-3899/24/7/002} {\bibfield  {journal}
  {\bibinfo  {journal} {J. Phys.}\ }\textbf {\bibinfo {volume} {G24}},\
  \bibinfo {pages} {1181} (\bibinfo {year} {1998})},\ \Eprint
  {http://arxiv.org/abs/hep-ph/9807358} {arXiv:hep-ph/9807358 [hep-ph]}
  \BibitemShut {NoStop}%
\bibitem [{\citenamefont {Diehl}(2003)}]{Diehl:2003ny}%
  \BibitemOpen
  \bibfield  {author} {\bibinfo {author} {\bibfnamefont {M.}~\bibnamefont
  {Diehl}},\ }\href {\doibase 10.1016/j.physrep.2003.08.002,
  10.3204/DESY-THESIS-2003-018} {\bibfield  {journal} {\bibinfo  {journal}
  {Phys. Rept.}\ }\textbf {\bibinfo {volume} {388}},\ \bibinfo {pages} {41}
  (\bibinfo {year} {2003})},\ \Eprint {http://arxiv.org/abs/hep-ph/0307382}
  {arXiv:hep-ph/0307382 [hep-ph]} \BibitemShut {NoStop}%
\bibitem [{\citenamefont {Miller}(2019)}]{Miller:2018ybm}%
  \BibitemOpen
  \bibfield  {author} {\bibinfo {author} {\bibfnamefont {G.~A.}\ \bibnamefont
  {Miller}},\ }\href {\doibase 10.1103/PhysRevC.99.035202} {\bibfield
  {journal} {\bibinfo  {journal} {Phys. Rev.}\ }\textbf {\bibinfo {volume}
  {C99}},\ \bibinfo {pages} {035202} (\bibinfo {year} {2019})},\ \Eprint
  {http://arxiv.org/abs/1812.02714} {arXiv:1812.02714 [nucl-th]} \BibitemShut
  {NoStop}%
\bibitem [{\citenamefont {Dirac}(1949)}]{Dirac:1949cp}%
  \BibitemOpen
  \bibfield  {author} {\bibinfo {author} {\bibfnamefont {P.~A.~M.}\
  \bibnamefont {Dirac}},\ }\href {\doibase 10.1103/RevModPhys.21.392}
  {\bibfield  {journal} {\bibinfo  {journal} {Rev. Mod. Phys.}\ }\textbf
  {\bibinfo {volume} {21}},\ \bibinfo {pages} {392} (\bibinfo {year}
  {1949})}\BibitemShut {NoStop}%
\bibitem [{\citenamefont {Brodsky}\ \emph {et~al.}(1998)\citenamefont
  {Brodsky}, \citenamefont {Pauli},\ and\ \citenamefont
  {Pinsky}}]{Brodsky:1997de}%
  \BibitemOpen
  \bibfield  {author} {\bibinfo {author} {\bibfnamefont {S.~J.}\ \bibnamefont
  {Brodsky}}, \bibinfo {author} {\bibfnamefont {H.-C.}\ \bibnamefont {Pauli}},
  \ and\ \bibinfo {author} {\bibfnamefont {S.~S.}\ \bibnamefont {Pinsky}},\
  }\href {\doibase 10.1016/S0370-1573(97)00089-6} {\bibfield  {journal}
  {\bibinfo  {journal} {Phys. Rept.}\ }\textbf {\bibinfo {volume} {301}},\
  \bibinfo {pages} {299} (\bibinfo {year} {1998})},\ \Eprint
  {http://arxiv.org/abs/hep-ph/9705477} {arXiv:hep-ph/9705477 [hep-ph]}
  \BibitemShut {NoStop}%
\bibitem [{\citenamefont {Miller}(2007)}]{Miller:2007uy}%
  \BibitemOpen
  \bibfield  {author} {\bibinfo {author} {\bibfnamefont {G.~A.}\ \bibnamefont
  {Miller}},\ }\href {\doibase 10.1103/PhysRevLett.99.112001} {\bibfield
  {journal} {\bibinfo  {journal} {Phys. Rev. Lett.}\ }\textbf {\bibinfo
  {volume} {99}},\ \bibinfo {pages} {112001} (\bibinfo {year} {2007})},\
  \Eprint {http://arxiv.org/abs/0705.2409} {arXiv:0705.2409 [nucl-th]}
  \BibitemShut {NoStop}%
\bibitem [{\citenamefont {Carlson}\ and\ \citenamefont
  {Vanderhaeghen}(2009)}]{Carlson:2008zc}%
  \BibitemOpen
  \bibfield  {author} {\bibinfo {author} {\bibfnamefont {C.~E.}\ \bibnamefont
  {Carlson}}\ and\ \bibinfo {author} {\bibfnamefont {M.}~\bibnamefont
  {Vanderhaeghen}},\ }\href {\doibase 10.1140/epja/i2009-10800-0} {\bibfield
  {journal} {\bibinfo  {journal} {Eur. Phys. J.}\ }\textbf {\bibinfo {volume}
  {A41}},\ \bibinfo {pages} {1} (\bibinfo {year} {2009})},\ \Eprint
  {http://arxiv.org/abs/0807.4537} {arXiv:0807.4537 [hep-ph]} \BibitemShut
  {NoStop}%
\bibitem [{\citenamefont {Hawking}\ and\ \citenamefont
  {Ellis}(2011)}]{Hawking:1973uf}%
  \BibitemOpen
  \bibfield  {author} {\bibinfo {author} {\bibfnamefont {S.~W.}\ \bibnamefont
  {Hawking}}\ and\ \bibinfo {author} {\bibfnamefont {G.~F.~R.}\ \bibnamefont
  {Ellis}},\ }\href {\doibase 10.1017/CBO9780511524646} {\emph {\bibinfo
  {title} {{The Large Scale Structure of Space-Time}}}},\ Cambridge Monographs
  on Mathematical Physics\ (\bibinfo  {publisher} {Cambridge University
  Press},\ \bibinfo {year} {2011})\BibitemShut {NoStop}%
\bibitem [{\citenamefont {Abidin}\ and\ \citenamefont
  {Carlson}(2008)}]{Abidin:2008ku}%
  \BibitemOpen
  \bibfield  {author} {\bibinfo {author} {\bibfnamefont {Z.}~\bibnamefont
  {Abidin}}\ and\ \bibinfo {author} {\bibfnamefont {C.~E.}\ \bibnamefont
  {Carlson}},\ }\href {\doibase 10.1103/PhysRevD.77.095007} {\bibfield
  {journal} {\bibinfo  {journal} {Phys. Rev.}\ }\textbf {\bibinfo {volume}
  {D77}},\ \bibinfo {pages} {095007} (\bibinfo {year} {2008})},\ \Eprint
  {http://arxiv.org/abs/0801.3839} {arXiv:0801.3839 [hep-ph]} \BibitemShut
  {NoStop}%
\bibitem [{\citenamefont {Cosyn}\ \emph {et~al.}(2018)\citenamefont {Cosyn},
  \citenamefont {Freese},\ and\ \citenamefont {Pire}}]{Cosyn:2018thq}%
  \BibitemOpen
  \bibfield  {author} {\bibinfo {author} {\bibfnamefont {W.}~\bibnamefont
  {Cosyn}}, \bibinfo {author} {\bibfnamefont {A.}~\bibnamefont {Freese}}, \
  and\ \bibinfo {author} {\bibfnamefont {B.}~\bibnamefont {Pire}},\ }\href@noop
  {} {\  (\bibinfo {year} {2018})},\ \Eprint {http://arxiv.org/abs/1812.01511}
  {arXiv:1812.01511 [hep-ph]} \BibitemShut {NoStop}%
\end{thebibliography}%

\end{document}